\documentclass[12pt]{article}
\usepackage{mathtext}
\usepackage{epsfig,amsmath,amssymb,stmaryrd,enumerate}
\newcommand{\bZ}{\mathbb Z}
\newcommand{\bR}{\mathbb R}
\newcommand{\bC}{\mathbb C}

\newcommand{\cD}{{\cal D}}
\newcommand{\cL}{{\cal L}}
\newcommand{\cP}{{\cal P}}

\newcommand{\dd}{\partial}
\newcommand{\bd}{\bar\partial}

\newcommand\beq{\begin{equation}}
\newcommand\eeq{\end{equation}}

\newcounter{one}
\newcounter{two}
\newcounter{eighteen}
\newcounter{nineteen}
\newcounter{twenty}
\newcommand{\webref}[1]{this text is availble on the web page of S.P.Novikov: http://www.mi.ras.ru/\~~\!\!snovikov/#1.pdf.}

\begin{document}
\begin{center}
{\LARGE On the Ground Level of Purely Magnetic Algebro-Geometric 2D Pauli
 Operator (spin 1/2)}

\vspace{7mm}

{\large P.Grinevich, \footnote{P.Grinevich is supported by the RFBR
grant 09-01-12148? and by  the Program of Russian Academy of
Sciences "Fundamental problems of nonlinear dynamics"},

Landau Institute for Theoretical Physics, Chernogolovka\\ E-mail
pgg@landau.ac.ru

A.Mironov\footnote{A.Mironov is supported by the RFBR grant
09-01-00598-a},

Sobolev Institute of Mathematics, Novosibirsk\\
E-mail: mironov@math.nsc.ru

\medskip

S.Novikov\footnote{S.Novikov is supported by the RFBR grant
08-01-00054-a} \footnote{All papers of S.P.Novikov cited in the
References are available on the web page
http://www.mi.ras.ru/\~~\!\!snovikov, click {\bf Publications.} }

University of Maryland, College Park;\\ Landau Institute for
Theoretical Physics, Chernogolovka, E-mail novikov@ipst.umd.edu }

\vspace{1cm}

\end{center}

\begin{abstract}
Full manifold of the complex Bloch-Floquet eigenfunctions is
investigated for the ground level of the purely magnetic 2D  Pauli
operators (equal to zero because of supersymmetry). Deep connection of it
 with the 2D analog of the
''Burgers Nonlinear Hierarchy''  plays fundamental role here. Everything
 is completely calculated for the broad class of
Algebro-Geometric operators found in this work for this case.
For the case of nonzero flux the ground states were found by
Aharonov-Casher (1979) for the rapidly decreasing fields, and by
Dubrovin-Novikov (1980) for the  periodic fields. No
Algebro-Geometric operators where known in the case of nonzero flux.
  For genus $g=1$ we
found  periodic operators with zero flux, singular magnetic
fields and Bohm-Aharonov phenomenon. {\bf Our arguments imply that the delta-term
really does not affect seriously the spectrum nearby of the ground state.}
For $g>1$ our theory requires to use
only algebraic curves with selected point leading to the solutions elliptic
in the variable $x$ for KdV and KP in order to get periodic magnetic fields.
  The algebro-geometric case of genus zero leads, in particular,  to the slowly decreasing
lump-like magnetic fields with especially interesting variety of
ground states in the Hilbert Space $\cL_2(\bR^2)$.

\end{abstract}

\section{Introduction. Magnetic Pauli Operator and factorizable Schrodinger Operators}

A nonrelativistic 2D Pauli operator for the charged particles with
spin=1/2 moving in electric and magnetic fields
$E_\alpha=\dd_{\alpha}U$, $A_{\alpha}$
 (under the  Lorentz gauge condition)   has a form
   (see \cite{LL}, let $e=1$, $m=1/2$, we neglect
the  universal constants $c$, $\hbar$ whose values are inessential
here )
 \beq
\label{eq:1} L^{P}=\sum\limits_{\alpha=1,2}(p_{\alpha})^2 +
B\sigma_3+ U, \ ip_\alpha=\dd_\alpha+iA_\alpha, \
\sum\limits_{\alpha=1}^2 \dd_\alpha A_\alpha = 0, \eeq
$\sigma_\alpha$ are the Pauli matrices
$$
\sigma_1=\left(\begin{array}{cc} 0 & 1 \\ 1 & 0 \end{array} \right), \
\sigma_2=\left(\begin{array}{cc} 0 & -i \\ i & 0 \end{array} \right), \
\sigma_3=\left(\begin{array}{cc} 1 & 0 \\ 0 & -1 \end{array} \right), \
\sigma_0=\left(\begin{array}{cc} 1 & 0 \\ 0 & 1 \end{array} \right).
$$

Let $U=0$ (i.e. electric field is equal to zero). The operator
$L^{P}$ is reduced to the direct sum of two scalar Schrodinger
operators. They are written in the ``factorized'' form (see
\cite{AC,DN,DN2}): \beq \label{eq:2} L^{P}=QQ^+\oplus Q^+Q = L_+
\oplus L_- \eeq (we neglect all constants unimportant for our
goals). Here $Q=\dd + A$, $Q^+=-(\bd - \bar A)$, $-\bar A=A^{(\bar
z)},$ $\bd A+ \dd \bar A= 2B$ is magnetic field, $\dd=\dd_x-i\dd_y$,
$\bd=\dd_x+i\dd_y $, $\dd\bd=\Delta=\dd_x^2+\dd_y^2$.

Magnetic field $B$ here is perpendicular to the oriented plane
$(x,y)$. Therefore, the magnetic field has a sign. For the rapidly
decreasing field $B$ at $x^2+y^2\rightarrow\infty$, the magnetic
flux $\{B\}$ is finite by definition. The Operator $L^{P}$ is
nonnegative. Therefore, the ground state energy is equal to  zero
$\varepsilon_0=0$ or positive
 $\varepsilon_0>0$. Let $|\{B\}|\geq 1$ (in natural
quantum units). Then the ground state subspace in the Hilbert Space
is a linear space of the dimension $[\{B\}]=m$ (see \cite{AC}). For
the periodic case and integer flux $\{B\}\in\bZ$, the linear
subspace of ground states is infinite-dimensional. It is isomorphic
to the Landau level in a homogeneous magnetic field (see
\cite{DN,DN2}). The higher levels are separated from the ground
state by the {\bf nonzero gap} $\Delta_B>0$. According to the
literature of the late 80-s, this operator admits a
``supersymmetry'' transformation $P:L_+\rightarrow L_-\rightarrow0$,
$P^2=0$, $P:\Psi\rightarrow Q_+\Psi$ for $\{B\}>0$, $\Psi\in L_+$;
All positive energy levels $\varepsilon>0$ are  degenerate  since
$(\Psi,P\Psi)$ both are eigenfunctions. For the zero energy we have
$P\Psi=0$, $\varepsilon_0=0$, if function $\Psi$ belongs to the
Hilbert space $\cL_2(\bR^2)$. Using the ancient language of
 \Roman{nineteen} (or even  \Roman{eighteen}) century, there exists a
``Laplace transformation'' of the

2D second order scalar operators (see \cite{NV1})
\begin{align}
& L=(\dd_x+A)(\dd_y+D)+U, \ U=e^f \nonumber \\
& L\rightarrow \tilde L=e^f (\dd_y+D) e^{-f} (\dd_x+A) +U \\
& \Psi\rightarrow \tilde\Psi = (\dd_y+D) \Psi \nonumber
\end{align}
The equality $L\Psi=0$  implies $\tilde L\tilde\Psi=0$. We have here
$L=Q_1Q_2+U$.

In the selfadjoint elliptic case, studied in detail in \cite{NV1} from
 the point of view of spectral properties of operators, we have
\begin{align}
& L=QQ^+ +U, \ U=e^f \nonumber \\
& Q=(\dd + A), \ Q^+=-(\bd - \bar A).
\end{align}
For the ``purely factorizable'' operators $U=\text{const}$, we have
$\tilde L=Q^+ Q + U$ (let $U=0$). The Laplace Transformation
$\Psi\rightarrow\tilde\Psi=Q^+\Psi$ coincides with the
``Supersymmetry'' with $P=Q^+$ in the sector $L^+$ and $P=0$ in the
second sector $L^-$. It acts on the whole spectrum
$L\Psi=\varepsilon\Psi$, $\tilde L\tilde\Psi=\varepsilon\tilde\Psi$.
The Ground States of $L^{P}$ are all ``instantons'', i.e. they
satisfy to the equation $Q^+\Psi=0$ if $\Psi\in\cL_2(\bR^2)$. Let
zero mode $L\Psi=0$ does not belong to the Hilbert space but
''belongs to the spectrum''. It simply means that its growth rate is
slower than some polynomial for $x^2+y^2\rightarrow\infty,$. The
 instanton argument disappears. In the last case the point $\varepsilon_0=0$ is
  the bottom of  continuous spectrum.
Even if no ``instanton'' type solutions  $\Psi$ of that kind exist,
  it is impossible to conclude immediately that the true ground state for the operator
$L$ is positive $\varepsilon_0>0$ (but it is highly probable).

In the case of nonzero flux $\{B\}\neq 0$ the ground energy of the
Pauli operator is equal to zero $\varepsilon_0=0$. For
$L^{P}=L_+\oplus L_-$ it is realized inside of the sector
 $L=L_+$ (if $\{B\}>0$) or $L=L_-$ (if $\{B\}<0$) (see  \cite{AC,DN,DN2}), in the rapidly decreasing
and periodic case (see also \cite{AS}, where other functional
classes of magnetic fields were considered). The rest of the
spectrum is twofold and separated from zero by a positive gap
$\Delta_B$. Interesting classes of the ``factorizable'' operators
$L$, having one more infinitely degenerate level except of the
ground one, were found in \cite{NV1}. These works have a ``soliton''
origin. Let us point out that the connection between Laplace
transformations and 2D Toda chain found in the soliton theory, was
in fact known in the \Roman{nineteen} century to Darboux and his
school. However,  all calculations at the end of \Roman{nineteen}
--- beginning of \Roman{twenty} centuries were purely formal, and
the elliptic case was completely missing.

In the present paper we investigate the algebrogeometric case. For
the smooth periodic operators (i.e vector-potentials are periodic)
we have {\bf magnetic flux equal to zero}, but for the degenerate
soliton-type case it might be not so.
 In our case the whole complex variety $\Gamma$ of the Bloch-Floquet
 zero level eigenfunctions
$L\Psi=0$ appears (''The Complex Fermi Curve''). This operator is
called ``algebro-geometric'' if genus is finite.

{\bf  A Purely Factorizable Reduction of the self-adjoint
Schrodinger operator $L=-(\dd+A)(\bd-\bar A)$ is studied here from
the point of view of algebro-geometric operators.} It was recently
found by the present authors \cite{GMN} using  the 2D  Soliton-Type
Completely Integrable System. The operator $L=\dd_x\dd_y+G\dd_y+S$
is by definition hyperbolic in the work \cite{GMN}. Its reduction
$S=0$ leads to very interesting ''2D Burgers Hierarchy'' which is
linearizable similar to the classical Burgers Equation. The spectral
meaning of this linearization is revealed in \cite{GMN}. For the
nonreduced system $L_t=(LH-HL)+fL$ the second operator
$H=\Delta+F\dd_y+A$ may have an interesting spectral theory in the
stationary ''finite-gap'' case $LH-HL=-f(x,y)L$. The operators $L$
and $H$ form ''The  Algebrogeometric Pair of PDE's Commmuting
Relative to the Level of Operator $L=0$'' according to the
terminology used by Krichever and Novikov in the late 1970s-early
1980s. The operator $H$  is elliptic here. Its study is the second
main goal of \cite{GMN}. It is easy to make $H$ smooth, periodic and
real. However, we failed to find nontrivial self-adjoint operators
$H$ within this approach. So the {\bf Conjecture} is formulated in
\cite{GMN}: {\bf For the smooth periodic self-adjoint 2D Schrodinger
operator in $R^2$ the Full Complex Manifold of  Bloch-Floquet
Eigenfunctions might contain Complex Algebraic Submanifolds only
belonging to one energy level (except some trivial cases which
essentially can be reduced to one-dimensional equation)}.

\section{Algebro-Geometric self-adjoint factorizable operators. The inverse spectral data}
As it was demonstrated above,  the purely magnetic 2D  Pauli
Operator $L^{P}=L_+\oplus L_-$ is a direct sum of two  ''Strongly
 Factorizable'' Schrodinger operators
\begin{align*}
& L_+=QQ^+ , \ L_-=Q^+Q \\
& Q=(\dd + A), \ Q^+=-(\bd - \bar A).
\end{align*}
Following \cite{DKN}, let us recall what is the algebro-geometric
operator with  periodic coefficients  $A$, $U$
$$
L=-(\dd + A)(\bd+D) +U,
$$
where $A$, $D$, $U$ are periodic in $x,y$.

Let us describe ''The Inverse Spectral Data'' for the Operator $L$:
we take nonsingular Riemann surface $\Gamma$ of genus $g>0$, two
marked ''infinite'' points $\infty_1$, $\infty_2$ with local
parameters $k'^{-1}(\infty_1)=0$, $k''^{-1}(\infty_2)=0$, and  set
of $g$ points ( a ``divisor'' of degree $g$)
$\cD=(\cP_1,\ldots,\cP_g)$. We write it as a formal sum
$\cD=\cP_1+\ldots+\cP_g$. In the work \cite{DKN} the ``Two-point
Baker-Akhiezer function'' was introduced, $\Psi(\cP,x,y)$,
$\cP\in\Gamma$, with the following properties:
\begin{enumerate}[a)]
\item It is meromorphic in the variable $\cP$ outside of infinities;
\item It has  following asymptotic near infinities $\infty_1$, $\infty_2$ :
\begin{align*}
\infty_1 &: \Psi=c(x,y) e^{k'\bar z}\left( 1+ O\left( \frac{1}{k'} \right) \right), \ \ \
& \left.\left.\frac{1}{k'}\right(\infty_1 \right)=0& \\
\infty_2 &: \Psi=    e^{k'' z}\left( 1+ O\left( \frac{1}{k''} \right) \right), \ \ \
& \left.\left.\frac{1}{k''}\right(\infty_2 \right)=0&.
\end{align*}
\item It has  poles of the first order in the points of divisor $\cD$ which are independent on the space variables
$x,y$.
\item $\Psi\equiv1$ at $x=0$, $y=0$.
\end{enumerate}
Such function satisfies to the equation
$$
L\Psi=0, \ L=\dd\bd-2(\ln c)_z\bd+U(x,y).
$$
Our function $\Psi$ is a Bloch-Floquet eigenfunction if $c(x,y)$,
$U(x,y)$ are periodic. In general, they are quasiperiodic but for a
dense set of Riemann surfaces $\Gamma$ these functions are periodic.
It is possible to write down $\Psi$, $c$, $U$ through the
$\Theta$-functions in a standard way of the periodic soliton theory
(see \cite{DKN} and surveys \cite{D}, \cite{DMN}). A Purely
Potential Reduction $c\equiv1$ was found for this data in
\cite{NV2}, \cite{NV3}. The Self-adjoint  reduction in the presence
of magnetic field
 $B=-\Delta(\ln c)/2=-\dd\bd(\ln c)/2\ne 0$ was found in \cite{Ch}:
 An Antiholomorphic Involution
 $\sigma:\Gamma\rightarrow\Gamma$, $\sigma^2=1$,
$\sigma(\infty_1)=\infty_2$ must exist such that \beq \label{eq:5}
\sigma(k')=-\overline{k''},\
\sigma(\cD)+\cD
\sim(K)+\infty_1+\infty_2, \eeq where $(K)$ is
divisor of zeros and poles of holomorphic 1-forms, and the symbol
$\sim$ means ``linear equivalence '' of divisors, i.e. every divisor
of zeros and poles of meromorphic function is equal to zero. These
conditions are sufficient and (probably) necessary, but no rigorous
proof of necessity was obtained in the literature yet.

Let us describe {\bf ``The inverse spectral data''} for our
''Factorizable Self-adjoint  Reduction''. It is  result of the
present work.

Riemann surface $\Gamma$ is degenerate
$$
\Gamma=\Gamma'\cup\Gamma'', \infty_1\in\Gamma', \ \infty_2\in\Gamma'',
$$
and the intersection $\Gamma'\cap\Gamma''$ is a set of $k+1$ points $Q'_0$,\ldots,$Q'_k$.
\begin{center}
\mbox{\epsfxsize=6cm \epsffile{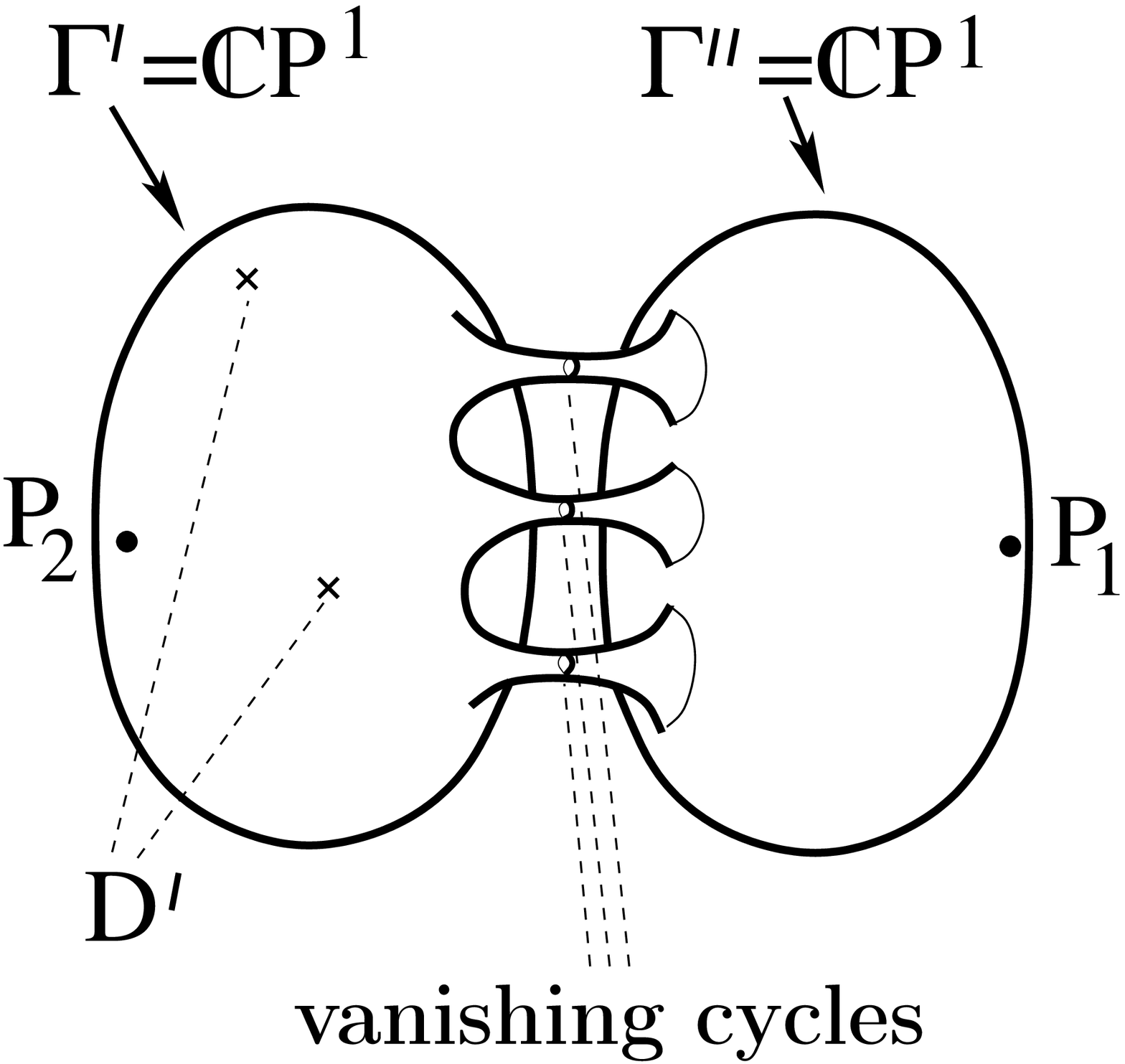}}
\mbox{\epsfxsize=6cm \epsffile{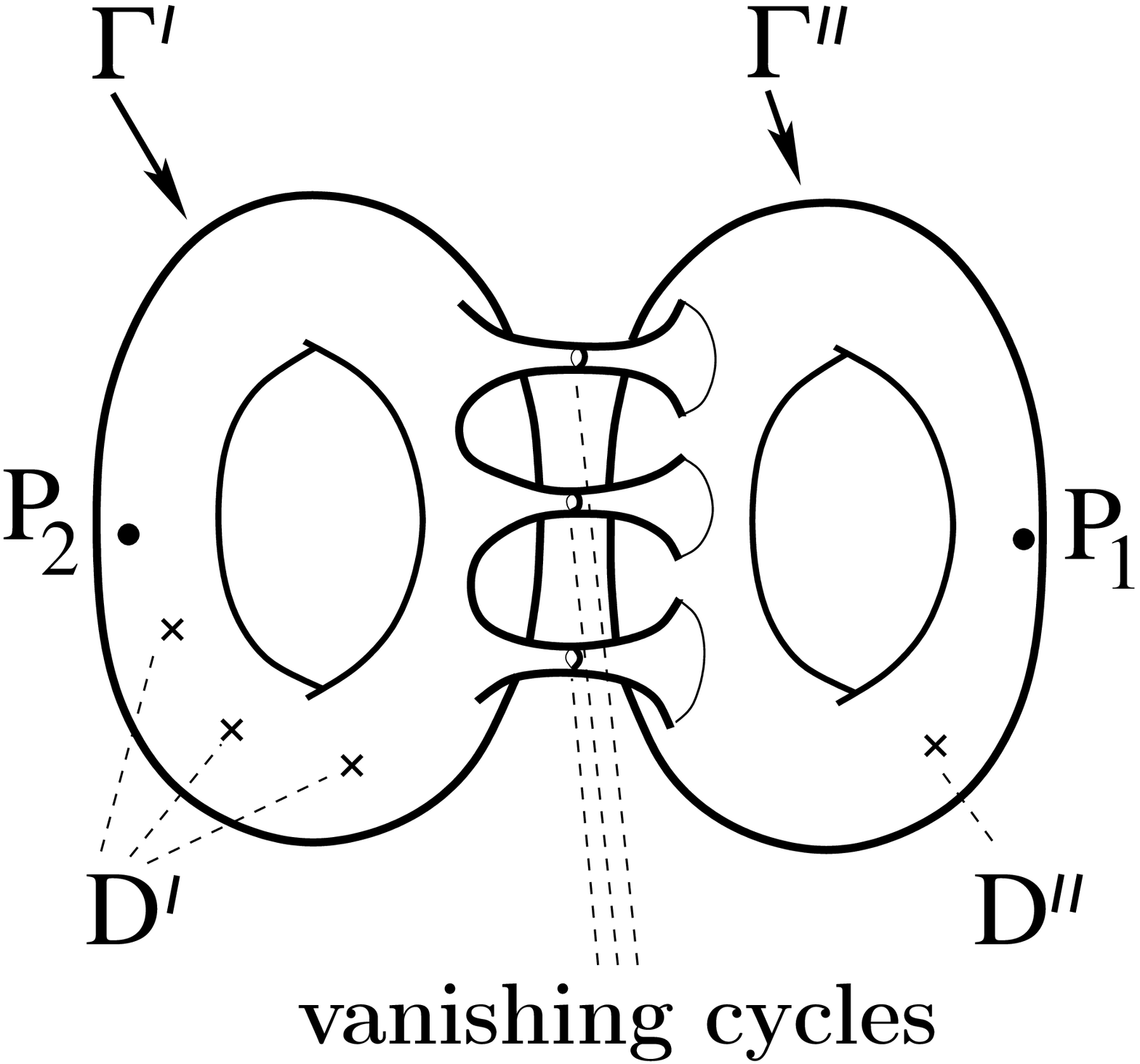}}

Fig. 1. $g=0$ \hspace{4cm} Fig. 2. $g=1$
\end{center}

An antiholomorphic involution should exist
$\sigma:\Gamma'\rightarrow\Gamma'', \ \Gamma''\rightarrow\Gamma'$,
which permutes points $\infty_1\xrightarrow[\sigma]{}\infty_2$,
$\infty_2\xrightarrow[\sigma]{}\infty_1$,
$(Q'_0,\ldots,Q'_k)\xrightarrow[\sigma]{}(Q'_{i_0},\ldots,Q'_{i_k})$.
Let $g=\text{genus}\ \Gamma'=\text{genus}\ \Gamma''$. Let us specify
$g+k$ points $\cD'=(\cP'_1,\ldots,\cP'_{g+k})$ on $\Gamma'$ and $g$
points $\cD''=(\cP''_1,\ldots,\cP''_{g})$ on $\Gamma''$ satisfying
to the linear equivalence: \beq \label{eq:6a}
\sigma(\cD'+\cD'')+\cD'+\cD''\sim(K)+\infty_1+\infty_2 \ \ \text{on}
\ \ \Gamma, \eeq where $(K)=(K')+(K'')$ is a divisor of 1-form
$\omega$ with conditions on residues:
\begin{align*}
\omega& = \omega' \ \ (\text{íà} \ \Gamma') \\
\omega& = \omega'' \ \ (\text{íà} \ \Gamma'') \\
-\underset{Q_j}{\rm Res} \ \ \omega' & = \underset{R_j=\sigma(Q_j)}{\rm
  Res} \omega'', \ \ \sigma(R)=Q.
\end{align*}

Let us reformulate it in terms of one curve $\Gamma'$ with local parameter $1/k'$,
set of points $Q_0$,\ldots,$Q_n$ and $\sigma(Q_j)=Q_{\sigma(j)}$.

Find $\Psi$-function on $\Gamma'$  such that:
\begin{enumerate}[1)]
\item It has poles of first order in points $\cD'$, ($g+k$ points).
\item It has asymptotic near $\infty_1$:
$$
\Psi=c(x,y) e^{k'\bar z}\left( 1+ O\left( \frac{1}{k'} \right) \right)
$$
\item $\Psi\raisebox{-2pt}{$\left|\rule{0pt}{10pt} \right.$}_{Q'_j}  =\varphi_j(z)$ holomorphic in $z$.
\end{enumerate} $(k+1)$ points $Q''_0=Q'_{i_0}$,\ldots,$Q''_k=Q'_{i_k}$ are fixed on the surface $\Gamma'$.
 Let us construct an antiholomorphic
(one-point) Baker--Akhiezer function $\varphi$   with properties:
$\varphi(\cP,z)$ is antiholomorphic in $\cP\in\Gamma'$; it has a
fixed divisor of first order poles $\sigma(\cD'')$ and asymptotic
$$
\varphi\sim e^{-\bar k' z}\left( 1+O\left( \frac{1}{\bar{k}'}\right)  \right), \ \varphi\equiv1
\ \ \text{ïðè} \ \ x=0, y=0.
$$

It is necessary to satisfy the condition
$$
\sigma(\cD'')+\cD' = (Q)+ (K')_{\Gamma'}+\infty_1
$$
 with restriction on residues (above) in the points $Q_j\sim
Q_{\sigma(j)}$ for the form defining $(K')$.

Let \beq \label{eq:6} \varphi_j(z)=\varphi(Q''_j,z), \ \
Q''_j=Q'_{\sigma(j)} \eeq We described everything in terms of one
nonsingular surface $\Gamma'$ because $\Gamma''=\sigma(\Gamma')$.

 Under these conditions we prove that our function $\Psi(\cP,z,\bar z)$ on the surface $\Gamma'\ni\cP$
satisfies to the equation below  in the space variables $z$, $\bar
z$:
$$
\tilde L_+ \Psi =0, \ \ \tilde L_+=(\dd+\tilde A)\bd,
$$
where $\tilde A=-\dd\ln{c}$. There is a constant $\alpha\ne0$ such
that the function $(\alpha c)$ is real. It generates symmetric
operator by the following
rule: let
\begin{align}
\label{eq:7}
& c=e^{2\Phi}, \ L_+=-e^{-\Phi} (\dd+A)\bd  e^{\Phi}=  \nonumber \\
& =-(\dd+A)(\bd-\bar A)=QQ^+, \ \ A=\tilde A/2   \\
& \Psi\Rightarrow e^{-\Phi}\Psi =\frac{1}{\sqrt{c}}\Psi  , \  1\Rightarrow \frac{1}{\sqrt{c}} \nonumber.
\end{align}
The operator $L_+$ is nonsingular, if there exists a constant
$\alpha$ such that $\alpha c(x,y)>0$. Let $\varepsilon_0=0$  be a
point of the spectrum for $L_+$. A natural  candidate for the ground
state is a function $1/\sqrt{c}$ satisfying to the equation \beq
\label{eq:8} Q^+ \left(1/\sqrt{c}\right)=(-\bd+\bar A) e^{-\Phi}=0.
\eeq As $(\dd+ A) e^{\Phi}=0$, the function $e^{\Phi}$ is also  a
candidate for the ground state for the operator $L_-=Q^+Q$. However
it is true only if these functions  belong to the spectrum in
$\cL_2(\bR^2)$. Let us formulate sufficient conditions for that:
\begin{enumerate}[a)]
\item The coefficients are periodic (quasiperiodic) and nonsingular  $\alpha c>0$. Both  functions
$\sqrt{\alpha c}$, $\sqrt{1/\alpha c}$ are positive and belong to
the spectrum. We know that $A=-(\dd\ln{c})/2$, $B=-(\Delta\
ln{c})/2$, and the magnetic flux is equal to zero:
$$
0=\iint\limits_{\oblong} B dz\wedge d\bar z =0, \ \ \text{where} \
\oblong \ \ \text{is an elementary cell}.
$$
It is true because $B=-(\dd\bd\ln{c})/2$, $Bdz\wedge d\bar z=
-\Phi_{z\bar z} dz\wedge d\bar z=-d(\Phi_{z}dz)= 1/2\cdot d(Adz)$,
so our flux  is an integral of the exact form.

\item Let us consider an exponential case now.

{\bf  Ground States of the Nonrelativistic Pauli operator:}

Let $L=L_+\oplus L_-$, $L_+=QQ^+$, $L_-=Q^+Q$, where $Q=(\dd+A)$, $Q^+=(-\bd+\bar A)$,
and $A=-2\Phi_z$, $\bar A=-2\Phi_{\bar z}$, where $\Phi$ is real.

In the whole class $c\rightarrow e^{W}c=c'$, where $W=\alpha x+
\beta y$ for real $\alpha$, $\beta$, we have the same magnetic field
$B=-\Delta\Phi=-(\Delta\ln{c})/2=-(\Delta\ln{(ce^{W})})/2$. The
operators $L_+$ and $L'_+$ are unitary equivalent. Indeed, we have:
\begin{align}
& L_{\pm} =  -(\dd_x-i\Phi_y)^2-(\dd_y+i\Phi_x)^2 \pm \Delta\Phi \\
& L'_{\pm} = -(\dd_x-i\Phi_y-i\beta/2)^2-(\dd_y+i\Phi_x+i\alpha/2)^2 \pm \Delta\Phi  \noindent
\end{align}
The Unitary Gauge Transformation
\begin{align}
& \Psi\rightarrow\Psi e^{i(-\beta x+\alpha y )/2} =\Psi'    \\
& L_+\rightarrow L'_+, \ L_-\rightarrow L'_-  \noindent
\end{align}
realizes this  equivalence. If any one of the functions $\sqrt{c'}$
or $1/\sqrt{c'}$ is bounded, then $\varepsilon_0=0$, and this
function serves the spectrum in $\cL_2(\bR^2)$. We have
$\Psi_+=e^{-i(\alpha x-\beta y)/2}\sqrt{c'}$, for $L_+$,
$\Psi_-=e^{-i(\alpha x-\beta y)/2}1/\sqrt{c'}$ for $L_-$. Thus we
constructed as many different  ground state vectors
$\varepsilon_0=0$ for $L_+$ or $L_-$ as  there are bounded functions
$\{e^Wc\}$ or $\{e^{-W}c^{-1}\}$ in this class.

{\bf Example 1.} Consider $c=1+e^y$. It does not depend on $x$. Here
$c^{-1}$ is bounded, and $c$ is  unbounded. Let
$c'_{\alpha\beta}=e^{\alpha x+\beta y}c$. To have bounded function
$1/c'$  in the class $c'=e^Wc, W=\alpha x+\beta y$, we need to
satisfy the conditions: $\alpha=0$, $-1\le\beta\le0$. So we obtain
{\bf continuum of eigenfunctions for the  ground energy level
$\varepsilon_0=0$} parametrized by the index $\beta$:
$$
\Psi_{\beta}=e^{i\beta x/2}\cdot 1/\sqrt{c'}, \ \beta\in[-1,0]
$$
 Here $c_{0,0}=c$, $\Psi_0=1/\sqrt{c}$. {\bf The unbounded functions
$(\sqrt{c},1/\sqrt{c})$ satisfy to the equation
 $L_+(\sqrt{c})=0$, $L_-(1/\sqrt{c})=0$ but do not belong to the spectrum.}

{\bf So we are coming to  the following conclusion:}

For the purely  exponential generating function $c>0$, the level
$\varepsilon_0=0$ is the lowest point of the spectrum if we can find
in the class $\{e^{\alpha x+\beta y}c\}$ for real $\alpha$, $\beta$
a bounded function $c'=e^{\alpha x+\beta y}c$ or $1/c'=e^{-\alpha
x-\beta y}/c$, $L^{P}=L_+\oplus L_-$. But it is always true for all
$k> 0$ where $k+1$ is the number of intersection points if all
coefficients $\kappa_j$ are positive.

In  the smooth periodic case we know that the
smooth periodic
functions
 $\sqrt{c}$ and $1/\sqrt{c}$ both are the ground states if $c$ is positive.
  They are periodic,
 and the zero energy level is always a twice degenerate point of the
 spectrum (not like in the case on nonzero flux).

In the next paragraphs we present calculations for the genuses
$g=0,1$.

\section{The Algebogeometric self-adjoint factorizable operators}

\subsection{Solutions of genus g=0.}

As we can see below, all algebrogeometric purely magnetic Pauli
operators with Complex Fermi Surface of genus zero correspond to the
functions $c$ of the form
$$c=\sum_j \kappa_je^{W_j}$$ Here $\kappa_j$ are constants, and
$W_j=a_jz+b_j\bar{z}$ are the linear forms with constant
coefficients. In general, all coefficients here are complex.
However,  for the selection of  physically meaningful self-adjoint
operators and real magnetic fields $B=-(\Delta\ln c)/2$ we are going
to formulate proper restrictions below. {\bf These ''Degenerate
Algebrogeometric Purely Magnetic Pauli Operators'' are the n Natural
Analogs of the ''Multisoliton Potentials'' for the 1D Schrodinger
Operators in the case of KdV.} It deserves to point out that our
generating functions $c$ are linear combinations of the elementary
exponents with constant coefficients. Indeed, the magnetic field $B$
is equal to $-1/2(\Delta\ln c)$, i.e. it is a strongly nonlinear
object. {\bf Such linear behavior of the quantity $c$ reflects the
main property of the ''2D Burgers Hierarhy'' discovered in the work
\cite{GMN}.} For KdV the Multisoliton functions are also constructed
as the second logarithmic derivatives of something which is indeed a
 nonlinear expression like some determinant made out of the
one-soliton functions. {\bf In spite of linearity of $c$, the
Spectral Theory is quite nontrivial for these 2D Purely Magnetic
Pauli Analogs of the Multisoliton Operators.}

 In the case of
genus zero our $\Psi$-function is written in the form
($k=k'\in\Gamma'$):
$$
 c\equiv u_0,
 \Psi=e^{k\bar{z}}\frac{u_0k^n+\dots+u_n}{(k-a_1)\dots(k-a_n)},\
 {\cal D}'=(a_1,\dots,a_n)\eqno{(11)}
$$
 (Here
 $n+1$ is the number of intersection points)
$$
 \Gamma=\Gamma'\cup\Gamma'',\ \Gamma'=\Gamma''=S^2
$$
with local parameters $k=k'(\Gamma')$ è $p=k''(\Gamma'')$; points
$\infty_1,\infty_2$ have the form $k=\infty (\infty_1)$, $p=\infty
(\infty_2)$. We have $\varphi(z,p)=e^{pz}$. The intersection points
are  $k_0,\dots, k_n$ for $\Gamma'$ and $p_0,\dots,p_n$ for
$\Gamma''$. They lead to the set of equations for $\Psi$ in these
points:
$$
 \Psi\vert_{k=k_j}=e^{p_jz}, j=0,\dots,n. \eqno{(12)}
$$
So our solution has a form:
$$
 c=\sum_{j=0}^n(-1)^je^{W_j(z,\bar{z})}\theta_j\frac{\Delta^{(n-1)}}{\Delta^{(n)}},
 W_j=p_jz-k_j\bar{z},\eqno{(13)}
$$
as  it follows from the system of linear equations

$$
 \left\{\begin{array}{c}
 u_0k_0^n+\dots+u_n=(k_0-a_1)\dots(k_0-a_n)e^{p_0z-k_0\bar{z}} \\
  \dots \ \dots\ \dots \\
 u_0k_n^n+\dots+u_n=(k_n-a_1)\dots(k_n-a_n)e^{p_nz-k_n\bar{z}}.
 \end{array}\right.\eqno{(14)}
$$
Here
$$
 \Delta^{(n)}=
 \left\Vert\begin{array}{ccc}
  k_0^n & \dots & 1 \\
  \dots &\dots&\dots \\
  k_n^n & \dots & 1
 \end{array}\right\Vert
 =\sqcap_{i<j}(k_i-k_j)
$$
and $\Delta_j^{(n-1)}$ are similar  Vandermonde determinants with
the set of generating numbers $(k_0,\dots,\hat{k}_j,\dots,k_n)$,
where $k_j$ is erased, $\theta_j=(k_j-a_1)\dots(k_j-a_n)$; $u_0=c$.

The quantity $c(x,y)=u_0$ is determined by the field $B$ up to the
transformation $c\rightarrow\alpha e^Wc=c',$ where $\alpha=const,$
$W=\gamma z+\delta\bar{z}$ because $B=-(\Delta \ln c)/2$ and
$-(\Delta\ln c')/2=-(\Delta\ln c)/2$,
$\Delta=\partial\bar{\partial}.$ Therefore we have exactly $n$
unknown coefficients in the formula (15):
$$
 j=0,\dots,n,\ c=\sum_{q=0}^{n} {\kappa}_qe^{W_q(z,\bar{z})}, \eqno{(15)}
$$
For the given  $k_j,p_j$ all coefficients ${\kappa_q}$ are
determined up to the common multiplier.

For the differentials below the conditions on residues should also
be satisfied:
$$
 \Omega_1=\frac{(k-a_1)\dots(k-a_n)dk}{(k-k_0)\dots(k-k_n)},
\ \Omega_2=\frac{s(p+\bar{a}_1)\dots(p+\bar{a}_n)dp}{(p-p_0)\dots(p-p_n)},
$$
where $s$ is a constant.
$$
  {\rm Res}_{k_j} \Omega_1+{\rm Res}_{p_j} \Omega_2=0.
$$

Choosing appropriate divisors ${\cal D}'=(a_1,\dots,a_n)$, we obtain
all  complex coefficients ${\kappa}_j\in{\mathbb C}$.

We need to classify such divisors  ${\cal D}'$ and linear forms
$W_j=p_jz-k_j\bar{z}$ that  $c$ is real and positive in the
equivalence class  $c\rightarrow\alpha c=c',$ $\alpha=const.$

For the reality of $c$, $(x,y)\in R$, $z=x+iy,\bar{z}=x-iy$,  our
expression must consists of the following terms:

1. The ''exponential type'' term for some index $j$:
$$
 \bar{p}_j=-k_j, W_j=p_jz+\overline{p_jz},
$$
${\kappa}_j$ is real and ${\kappa}_je^{W_j}$ is also real (a purely
real exponent)

2. The ''mixed type'' term for the pair of indices $(j,l)$:
$$
 p_l=-\bar{k}_j, k_l=-\bar{p}_j,
$$
$
 \bar{{\kappa}}_j={\kappa}_l, \ {\kappa}_je^{W_j}+{\kappa}_le^{W_l}={\kappa}_je^{W_j}+\bar{{\kappa}}_le^{\bar{W}_l}
$ is real. We assume that all points $k_j\ne k_q$, $j\ne q$ and
$p_j\ne p_q$, $j\ne q$ are distinct.

{\bf For the case 1:} We obtain  terms like  real exponent
${\kappa_j}e^{(\alpha_j x+\beta_j y)},$ where
$p_j=\alpha_j+i\beta_j,k_j=-\bar{p}_j,\
{\kappa_j}-\bar{\kappa_j}=0.$

{\bf For the case 2:} We obtain  terms of the form
$$
 e^{W_{R,j}}({\kappa}_j'\cos W_{I,j}-{\kappa}_j''\sin W_{I,j}), {\kappa}_j={\kappa}_j'+i {\kappa}''_j,
$$
$$
 W_j=W_{R,j}+i W_{I,j}=[(\alpha_j-\gamma_j)x-(\beta_j-\delta_j)y]+i[(\beta_j-\delta_j)x+(\alpha_j+\gamma_j)y],
$$
where $p_j=-\bar{k}_q=\alpha_j+i\beta_j, k_j=-\bar{p}_q=\gamma_j+i\delta_j.$

For $W_{I,j}=0$, we have the case 1: $p_j=-\bar{k}.$

3. The ''purely trigonometric type'' appears as another special
subcase of the case 2 if $W_{R,j}=0$ or $\alpha_j=\gamma_j,
\beta_j=-\delta_j,$ i.e. $k_j=\bar{p}_j,$ $W_{I,j}=(\beta_j
x+\alpha_j y), e^{W_j}=e^{p_jz-\bar{p}_j\bar{z}}.$

In all these cases $c$ is real. The mixed case 2 leads to the zeros
of
 $c$ and   singularities of magnetic field if they
are not `` blocked'' by other stronger terms.

$$
 c=\sum_{j} {\kappa}_je^{W_j(x,y)}, \ c\rightarrow{\kappa} e^Wc=c',
$$
where all $\kappa_j$ are real. Let $c=c^++c^-$. Here $\kappa_j>0$
for $j\in{\rm\Roman{one}}$, $\kappa_j<0$  $j\in{\rm\Roman{two}}$.

Consider at first the case  $c=c^+$, i.e. $\kappa_j>0$ for all $j$.
So we have $c>0$. As one can  see (below),  the magnetic field
$B=-(\Delta \ln c)/2$ is bounded in ${\mathbb R}^2$. In the class
$\{\kappa e^Wc\}$ both $\sqrt{c'}, \frac{1}{\sqrt{c'}}$ never can be
bounded. Either they both exponentially increase along some
directions at $x^2+y^2\rightarrow\infty$ or one of them (i.e.
($1/\sqrt{c'}$)) is bounded. In the last case the pair $\{c,W\}$ or
simply a function $e^Wc=c'$ defines the ground state. As we can see
below, such functions $c'$ form a domain inside of  the convex
polygon $T$ in $R^2$ which is always nonempty. This domain is
completely determined by the set of linear forms $W_j$ in the class
$\{W_j\}$: it is a convex hull of the set of points
$(\alpha_j,\beta_j)\in R^2, W_j=\alpha_jx+\beta_jy$ (see below).

2. Let us consider {\bf the
 purely trigonometric case}. Here we have the
cases of odd and even numbers of intersection points $n+1$. They are
drastically different.

 a)The number of intersection points $n+1$ is even.

b)The number of intersection points $n+1$ is odd.

 $$ a) \
c=\sum_{j=0}^{\frac{(1+n)}{2}}{\kappa}_j'\cos
W_{I,j}+{\kappa}_j''\sin W_{I,j}, \\
$$
where ${\kappa}_j={\kappa}_j'+i{\kappa}_j'',
W_j=iW_{I,j}=-k_j\bar{z}-\bar{k}_jz, k_j=\alpha_j+i\beta_j$. Here
function $c$ always has zeros.
$$
b) \ c=1+\sum_{j=1}^{\frac{n}{2}}{\kappa}_j'\cos
W_{I,j}+{\kappa}_j''\sin W_{I,j}, \\
$$
For the appropriate constants $\kappa', \kappa''\in\bR$ we have
$c>0$ and magnetic field $B=-(\Delta\ln c)/2$ is smooth, periodic
and has  zero flux through the elementary cell of periodic lattice
(or the quasiperiodic mean value if $c$ is quasiperiodic). It would
be interesting to describe corresponding domains in the space of
constants.
  The set of lines
$W_{I,j}=\alpha_jx+\beta_jy$, should pass through the integer
vectors of the lattice in ${\mathbb R}^2.$  Otherwise, the fields
are quasiperiodic. Our {\bf conclusion} is that in the regular
trigonometric case both functions $\sqrt{c}$, $1/\sqrt{c}$ {\bf are
periodic and positive;  they are the ground states in both sectors
$L_+,L_-$} of the Operator $L^P$. In the general quasiperiodic case
the situation is the same.

{\bf  A Curious Remark.} There are ``critical'' values of constants
$\kappa'_j$ $\kappa''_j$ such that $c$ has  isolated zeroes $c=0$
(repeated periodically). It is possible to choose  parameters
 such that  we have in this critical point an isotropic hessian
$d^2c=\pm a^2(dx^2+dy^2)$. Then the magnetic field has a
$\delta$-shaped singularity
$B=-(\Delta\ln{c})/2\sim\delta(x-x_0,y-y_0)$.

{\bf Example 2.} Let $n=4$. We demonstrate here a simplest
nonsingular purely trigonometric (i.e degenerate algebrogeometric)
operator, essentially dependent on both variables $x,y$: we write
$\Psi$-function  in the form
$$
  \Psi=e^{k\bar{z}}\frac{u_0k^4+u_1k^3+u_2k^2+u_3k+u_4}{(k^2-a_1^2)(k^2-a_2^2)}, \ {\cal D}'=(a_1,a_2,-a_1,-a_2),
$$
and $\varphi=e^{pz}.$

 Take the intersection points of $\Gamma'$ and
$\Gamma''$ in the form $0,k_1,k_2,-k_1,-k_2$ for $\Gamma'$ and
$0,p_1,p_2,-p_1,-p_2$ for  $\Gamma''$. Let
$$
 p_1=k_1\in{\mathbb R},\ p_2=-k_2=iK\in i {\mathbb R}.\eqno{(16)}
$$
In this case the antiinvolution $\sigma:k\rightarrow -\bar{p}$ is
correctly defined on $\Gamma$. The differentials look like
$$
 \Omega_1=\frac{(k^2-a_1^2)(k^2-a_2^2)dk}{(k^2-k_1^2)(k^2-k_2^2)k},
$$
$$
 \Omega_2=\frac{s(p^2-\bar{a}_1^2)(p^2-\bar{a}_2^2)dp}{(p^2-p_1^2)(p^2-p_2^2)p},
$$
where $s$ is some number.  The condition on the residues
$$
 {\rm Res}_0 \Omega_1+{\rm Res}_0 \Omega_2=0,\ {\rm Res}_{\pm k_j} \Omega_1+{\rm Res}_{\pm p_j} \Omega_2=0.
$$
must be valid. In the points of intersection we have
$$
 \Psi(0)=1,
 \Psi(k_1)=e^{p_1z},\Psi(k_2)=e^{p_2z},\Psi(-k_1)=e^{-p_1z},\Psi(-k_2)=e^{-p_2z}.
$$
 So the equalities  follow:
$$
 u_4=a_1^2a_2^2,
$$
$$
 u_0k_1^4+u_1k_1^3+u_2k_1^2+u_3k_1=-a_1^2a_2^2+(k_1^2-a_1^2)(k_1^2-a_2^2)e^{p_1z-k_1\bar{z}},
$$
$$
 u_0k_2^4+u_1k_2^3+u_2k_2^2+u_3k_2=-a_1^2a_2^2+(k_2^2-a_1^2)(k_2^2-a_2^2)e^{p_2z-k_2\bar{z}},
$$
$$
 u_0k_1^4-u_1k_1^3+u_2k_1^2-u_3k_1=-a_1^2a_2^2+(k_1^2-a_1^2)(k_1^2-a_2^2)e^{-p_1z+k_1\bar{z}},
$$
$$
 u_0k_2^4-u_1k_2^3+u_2k_2^2-u_3k_2=-a_1^2a_2^2+(k_2^2-a_1^2)(k_2^2-a_2^2)e^{-p_2z+k_2\bar{z}}.
$$
Sum of the second equality with the fourth equality, and of the
third equality with the fifth one are written below:
$$
 2u_0k_1^4+2u_2k_1^2=-2a_1^2a_2^2+(k_1^2-a_1^2)(k_1^2-a_2^2)2\cos(2k_1y),
$$
$$
 2u_0K^4-2u_2K^2=-2a_1^2a_2^2+(K^2+a_1^2)(K^2+a_2^2)2\cos(2Kx).
$$
Take
$$
 a_1\in{\mathbb R},a_2=i a\in i{\mathbb R}.
$$
Using (16)  and taking $s=-1$, we can see that the conditions on
 residues of the differentials  $\Omega_1$ and $\Omega_2$ are
satisfied. We have
$$
 c=u_0=\frac{a_1^2a^2}{k_1^2K^2}\left(1-A\cos(2k_1y)- B\cos(2Kx)\right),
$$
where
$$A=\frac{(k_1^2-a_1^2)(k_1^2+a^2)}{k_1^2(k_1^2+K^2)},\ B=\frac{(K^2+a_1^2)(K^2-a^2)}{K^2(k_1^2+K^2)}.$$
For $K=10,\ k_1=5, a_1=2, \ a=1$ we obtain
$$
 A=\frac{546}{3125},\ B=\frac{2574}{3125},\ A+B=\frac{624}{625},
$$
So the magnetic field is smooth and periodic.

{\bf Example 3.} Consider now {\bf the Full Class of the Purely
Exponential Real Type functions $C$}.  We introduce below an
important notion of ''the Indicator of Growth'' for the set
$\{W_j\}$ of all real exponents entering formula for $c$ with
positive coefficients. Suppose this  function $c$ and therefore
$\sqrt{c}$ grows exponentially in all directions. This is a Stable
Property. The function $\frac{1}{\sqrt{c}}$ has exponential decay in
${\mathbb R}^2$. Many other functions $1/\sqrt{c'}$ in the same
class $c'=e^Wc$ are such that $1/\sqrt{c'}$ automatically have
similar decay (for example, it is true for all ``small'' linear
forms $W=\varepsilon (ax+by),\varepsilon \rightarrow 0$).

{\bf Consider first the Unstable Case $n=1$, i.e. with 2
intersection points}. Let us take absolutely typical example
$c=1+e^y$ following notations in the Example 1 above. The polygon
$T$ numerating all bounded functions $1/c'$ in the class $c'=e^Wc$,
coincides with a segment $\beta\in [-1,0],W=\alpha x+\beta y$. It
does not have inner points. We never have $c'$ in this class which
has exponential growth in all directions. Magnetic field here
depends on one variable. {\bf The same result is true for all cases
with $n+1=2$ where $c=\kappa_1 e^{W_1}+\kappa_2 e^{W_2}$}. {\bf The
Unstable cases for all $n>1$ are given by the sets linear forms
$\{W_j\}$ such that all differences are proportional to one linear
form with constant coefficients. Magnetic field here depends on one
variable only.}

\begin{center}
\mbox{\epsfxsize=5cm \epsffile{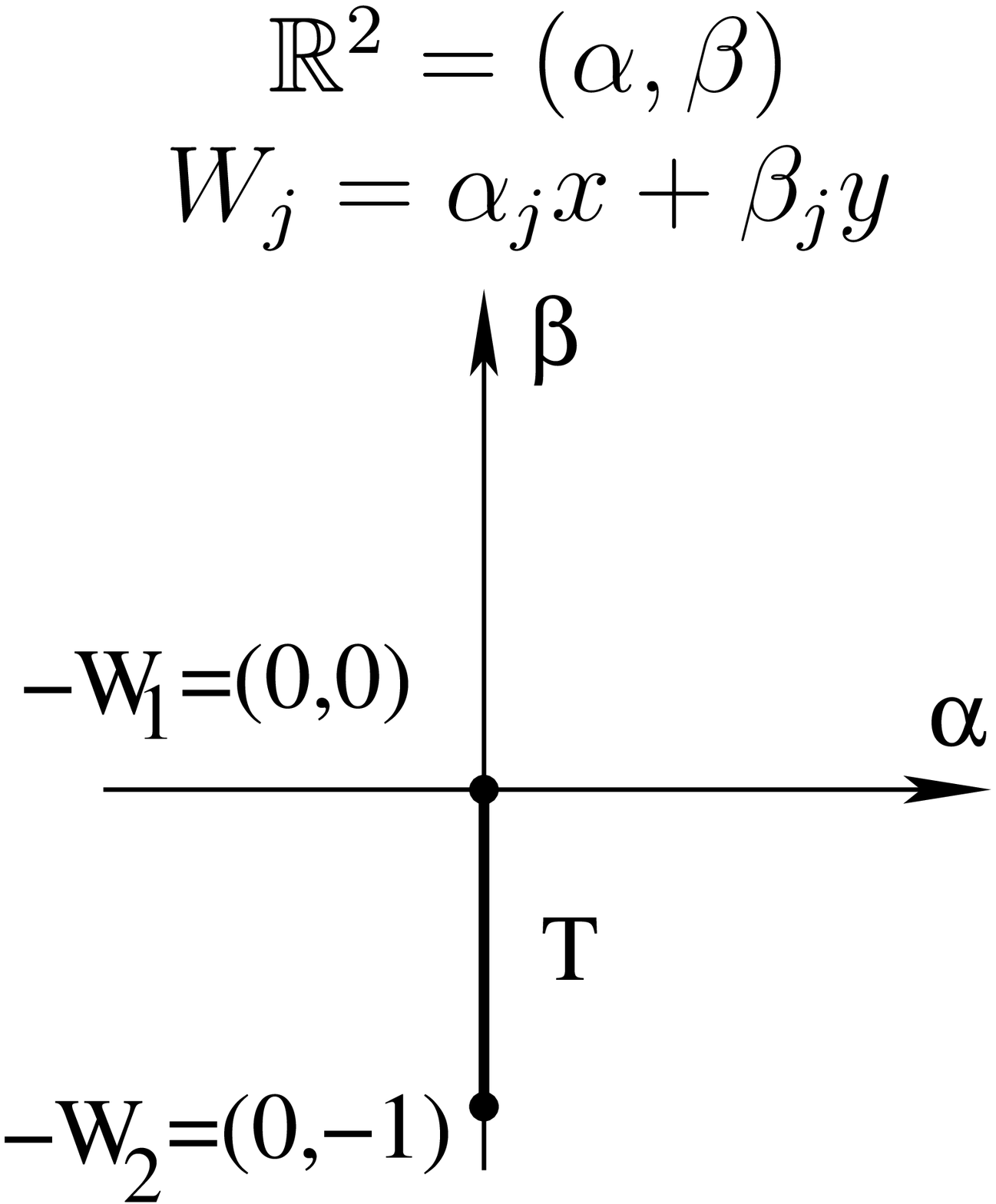}}

Fig. 3
\end{center}

Let us consider  examples of the indicators of growth in this class.
Choose $W=e^{-y/2}$. We have $c'=e^{y/2}+c^{-y/2}$. The indicator of
growth has zeroes $I_{W'_j}(\varphi)=0$ exactly in two points
$\varphi=0,\pi$. Put $W=e^{(x-y)/2}$.  We get
$c''=e^{(x+y)/2}+c^{(x-y)/2}$. For this case  $I_{W''_j}(\varphi)=0$
on the connected segment. However, the zero set of the indicator of
growth is never empty for $n=1$. It is not surprising because
magnetic field always depends on one variable for $n=1$.

\begin{center}
\mbox{\epsfxsize=6cm \epsffile{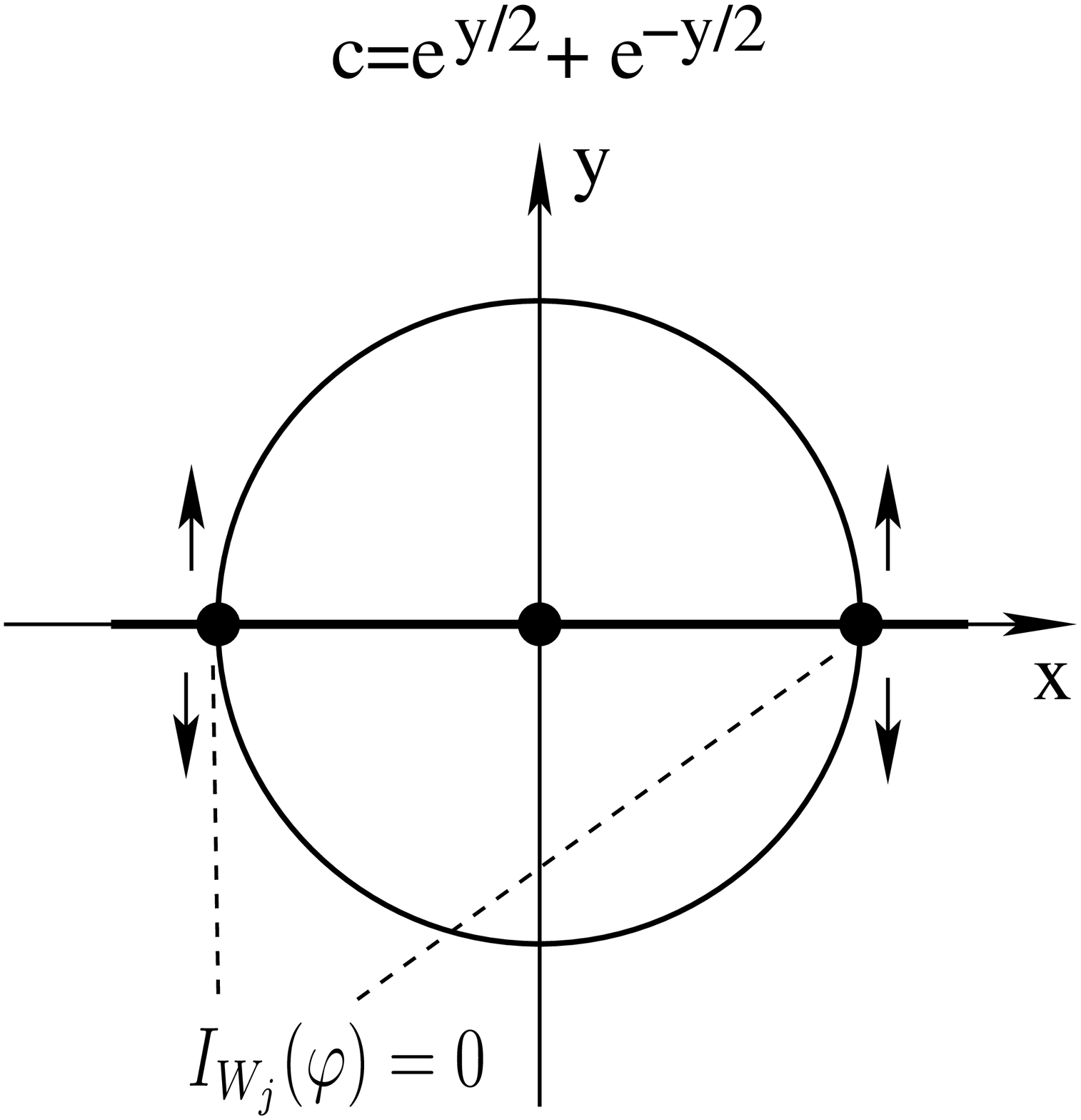}} \mbox{\epsfxsize=6cm
\epsffile{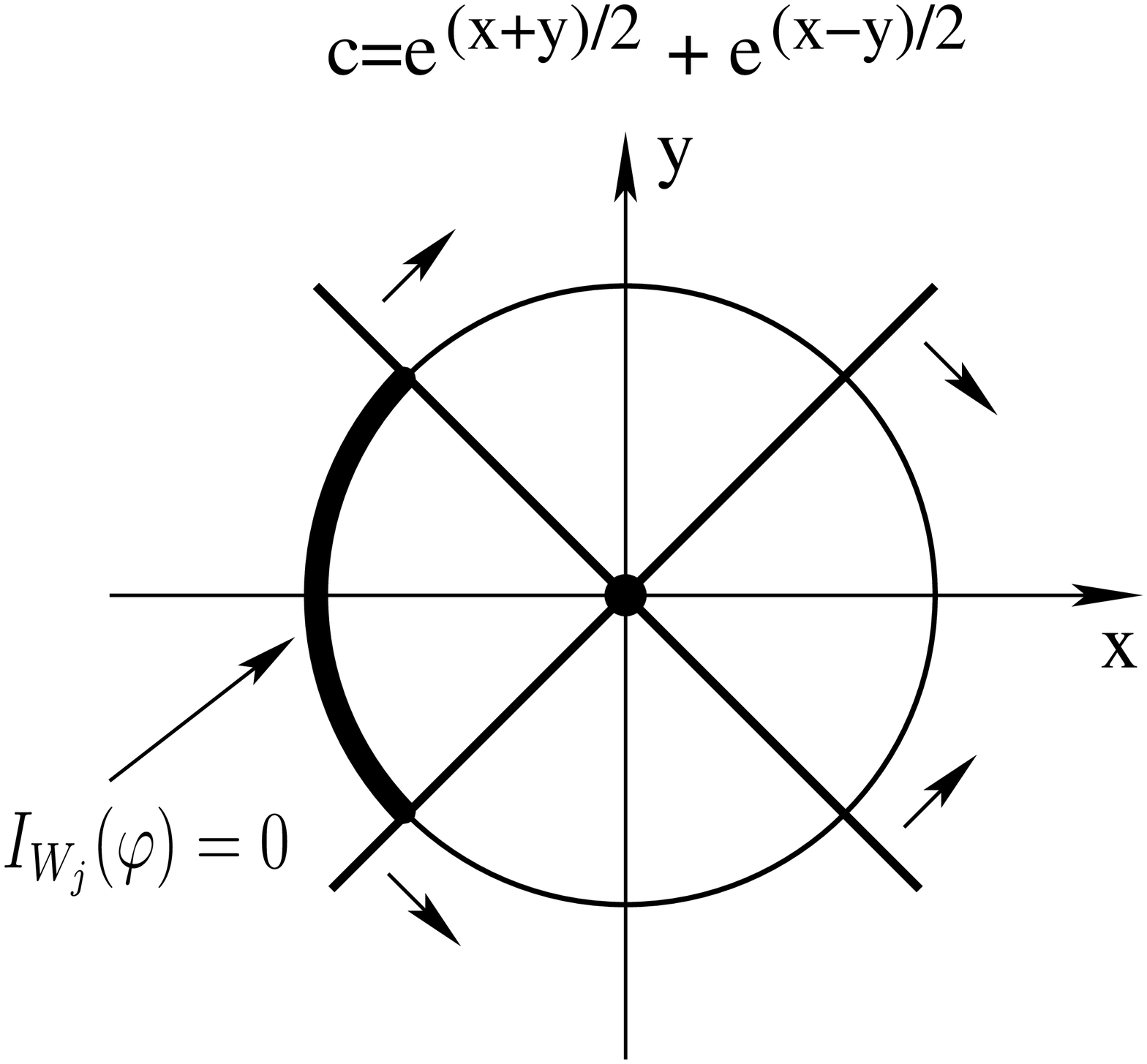}}

Fig. 4 a. \hspace{5cm} Fig. 4 b.
\end{center}
(the growth directions for  $W_j$ are shown by the rows; every ray
from the zero point belongs to at least one sector of exponential
growth provided by the 3 exponents entering $c$.)

The Stable Cases start with $n=3$ (like in Fig 4,5). Let, for
example,
$$c=e^{W_1}+e^{W_2}+e^{W_3}=e^x+e^y+e^{-x-y}$$

\begin{center}
\mbox{\epsfxsize=7cm \epsffile{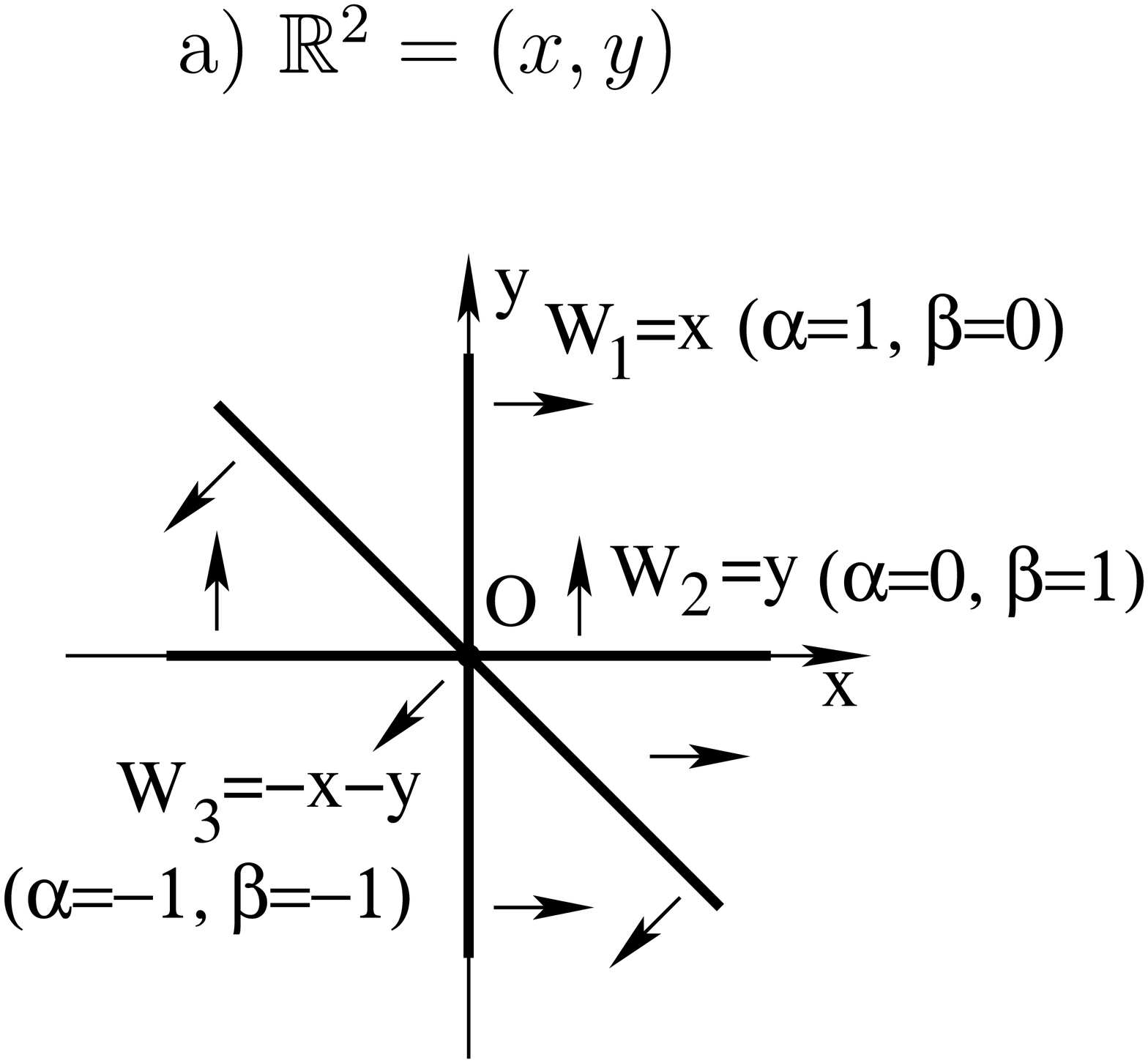}} \mbox{\epsfxsize=6cm
\epsffile{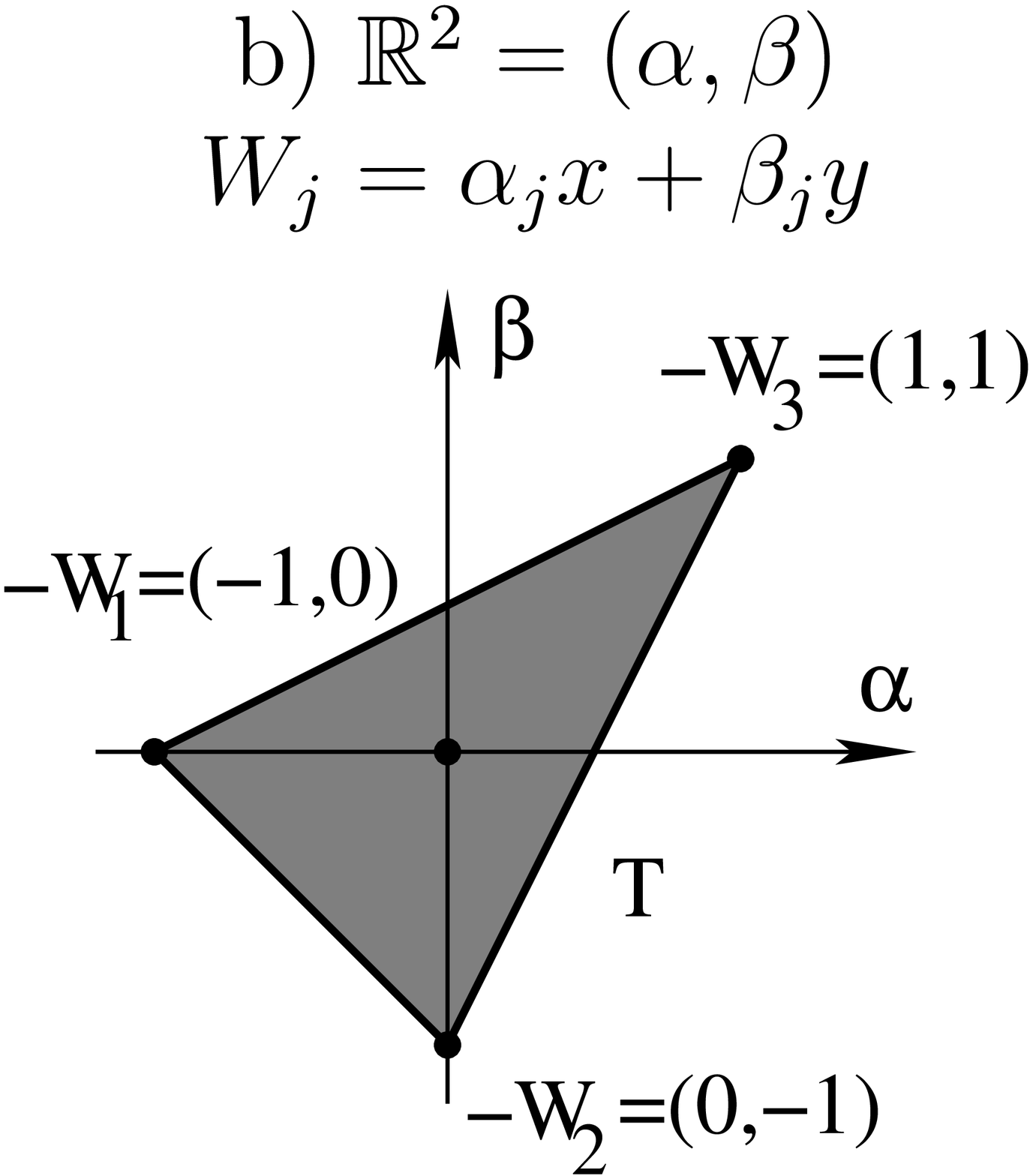}}

Fig. 5 a. \hspace{5cm} Fig. 5 b.
\end{center}

In this example the function $\frac{1}{c'}=\frac{1}{e^{W}c}$ is
bounded  if and only if $W\in T$\, where $T$ is  a triangle with
vertices $(-W_1,-W_2,-W_3)\subset\bR^2$ with coordinates
$(\alpha,\beta)$. It belongs to the Hilbert Space (i.e. it is square
integrable function on the $x,y$-plane $R^2$) if and only if
$c'=e^Wc$ where $W$ belongs to the interior of $T$.

{\bf We describe below all nonsingular cases for the exponential
generating functions $c$:}

{\bf Definition.} Let $W=\alpha x+\beta y$ be a real linear form. We
call the function on the circle $I_W\geq 0$,
$I_W(\varphi)=\max(\alpha \cos\varphi+\beta\sin\varphi, 0)$  {\bf
``the Indicator of Growth``}  for the linear form $W=\alpha x+\beta
y$. For the set of real linear forms $\{ W_j\}$ we call function
$I_{\{W_j\}}(\varphi)=\underset{j}\max(I_{W_j})\geq 0$ the {\bf
``the Indicator of Growth of this set``}. It is what some people
call ''a Tropical Sum''.

\noindent I. Consider the case
$$
c_{\rm\Roman{one}}=\sum_je^{W_j}{\kappa}_j, \ {\kappa}_j>0 \
\text{or} \ c_{\rm\Roman{two}}=\sum_je^{W_j}{\kappa}_j, \
{\kappa}_j<0.
$$

{\bf Let us note that the indicators of growth are different for the
different sets within the same class  $\{ W_j \}$ and $\{ W_j+W\}$.
 It is an invariant of the set of exponents entering the function
$c$, not of magnetic field. It does not depend also on the
coefficients $\kappa_j$ entering the function $c$.}

There are  following possibilities:

1. $I_{\{W_j\}}(\varphi)>0$ for all $\varphi\in S^1$ (see Fig.~6 a).

2.  $I_{\{W_j\}}(\varphi)=0$ on the connected closed segment on the
circle $\varphi\in S^1$ (see Fig.~6~b).

3. $I_{\{W_j\}}(\varphi)=0$ in the isolated points
$\varphi=\varphi_1,\dots,\varphi_s\in S^1$. Then either  $s=1$  or
$s=2$ (see. Fig.~4~a to the Example 2).
\begin{center}
\mbox{\epsfxsize=6cm \epsffile{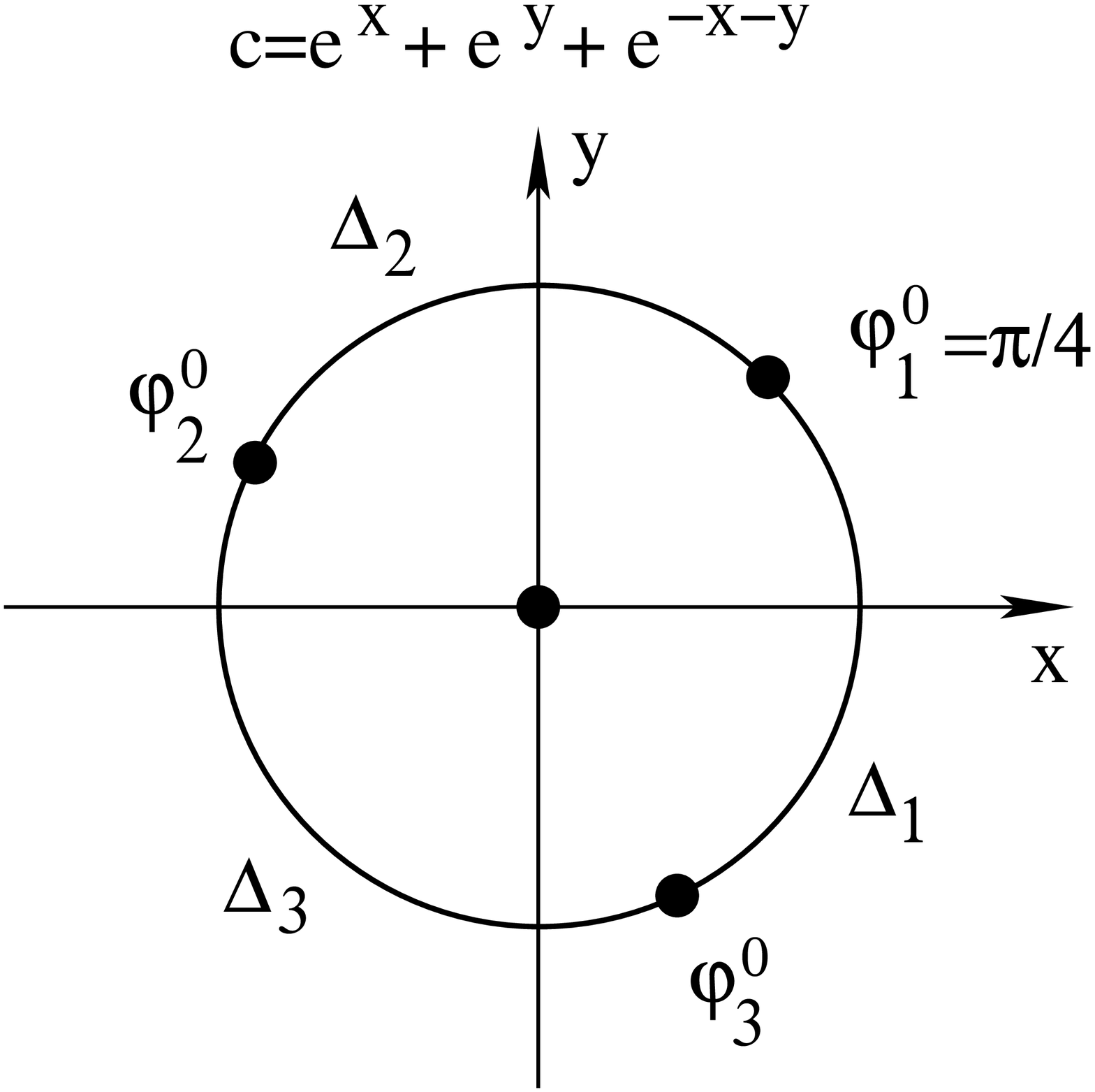}}
\mbox{\epsfxsize=6cm \epsffile{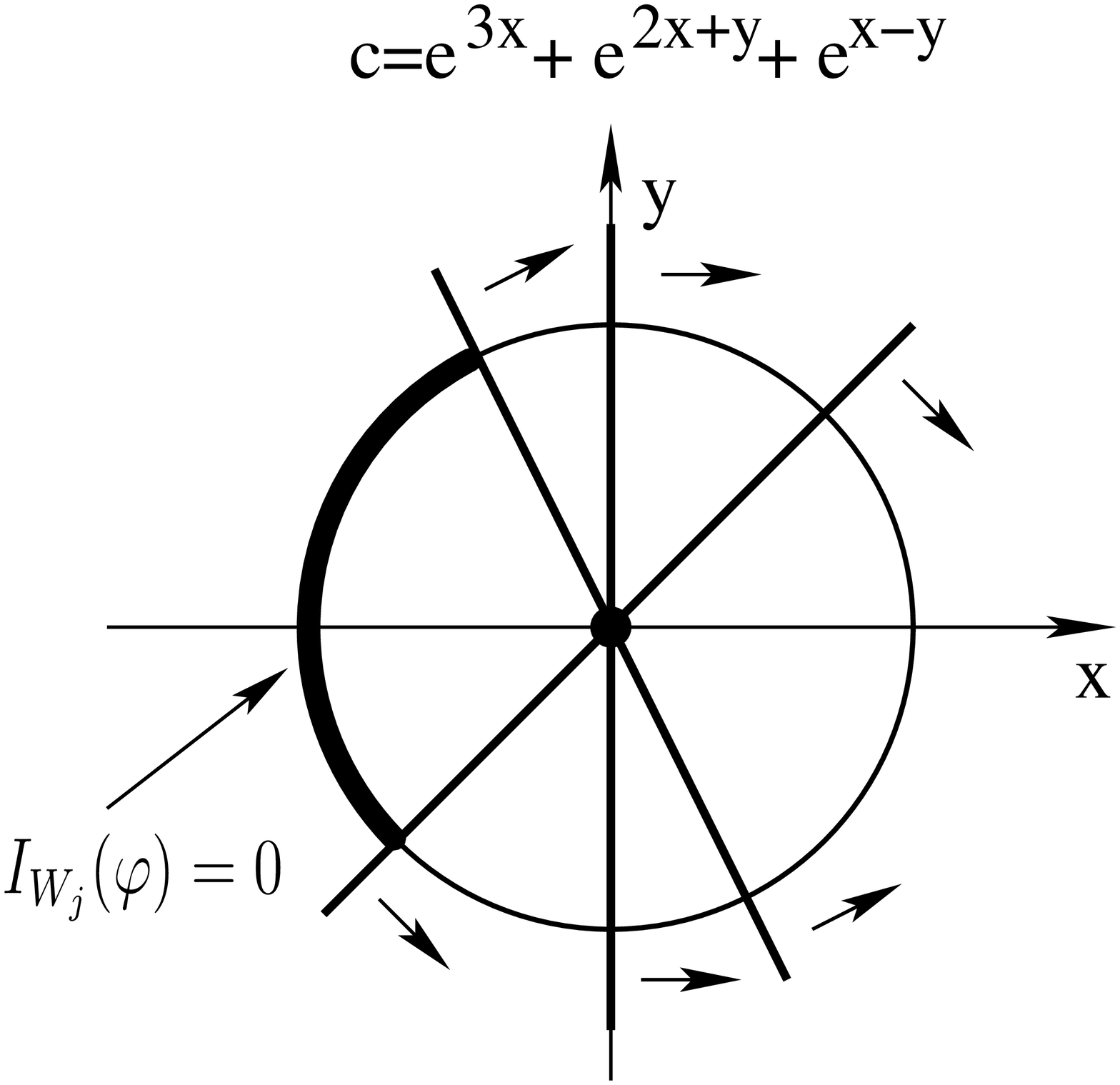}}\\
\parbox{6cm}{\center $\cos(\varphi^0_1)=\sin(\varphi^0_1)=\frac{1}{\sqrt{2}}$ \\
$\cos(\varphi^0_2)+2\sin(\varphi^0_2)=0$ \\
$2\cos(\varphi^0_3)+\sin(\varphi^0_3)=0$ \\
$I_{\{W_j\}}(\varphi)>0$\\ \vspace{1mm} Fig. 6 a.}
\parbox{6cm}{\center Zones of positivity for $I_{\{W_j\}}(\varphi): \ \ \ \left\{\begin{array}{l} x\ge0\\ y\ge -2x \\
  y\le x \end{array}\right.$\\ \vspace{1mm}  Fig. 6 b.}

\end{center}

We choose in the class $\{ce^W\}$ a representative $c'$ with minimal
set of zeroes of the indicator of growth $W_j\rightarrow W_j+W=W_j'$
for $c'=e^Wc$.  Possibility 3 with 2 opposite zeroes
$I_{W_j}(\varphi_1)=I_{W_j}(\varphi+\pi)$ is realized for
$c=e^{W_1}+e^{-W_1}$. It is impossible to reduce it to the case 1.
If $s=1$, then it is possible to reduce it to the ''Stable Positive
Case'' case $I_{W_j+W}(\varphi)>0$ by choosing $W$. The last case we
call ``stable''. The function $c'$ in this class $\{W_j+W=W'_j\}$
where $I_{W'_j}(\varphi_1)=0$ has an isolated zero,  we call  ``a
boundary function''. It is easy to see that in the stable case where
the function $c>0$ grows exponentially in all directions, its
opposite $c^{-1}$ has exponential decay in all directions.

Let $c'=c_0\in\{ce^W\}$ be stable (i.e has exponential growth in all
directions). All such  functions $c_0^{-1/2}$ are the ground states
for $L_{-}$ where $L^P=L_{+}\oplus L_{-}$. All  stable functions
$c''=c_W$ from the class $\{c_0e^W\}$ define the ground states for
$L^P$ located in the sector $L_-$.
$$
 \Psi_W= \frac{e^{-i(\alpha y-\beta x)/2}}{\sqrt{c_0}e^{(\alpha x+\beta
 y)/2}}=$$
 $$
 =\frac{e^{-i(\alpha y-\beta x)/2}}{\sqrt{c_W}}.
$$
{\bf Conclusion.} The operator $L_{-}$ has the family of the square
integrable ground states $\Psi_W$, parametrized by the linear forms
$W=\alpha x+\beta y$ (or by the pair $\alpha,\beta)\in R^2$ such
that the set $\{W_j+W\}$  has strictly positive indicator of growth
$I_{W_j+W}>0$ everywhere on the circle. This domain $T$ on the plane
${\mathbb R}^2$ with coordinates $(\alpha,\beta)$ is a {\bf convex
polygon} $T$. The interior of $T$ is nonempty for all stable cases
(i.e. if the set of linear forms $W_j-W_s$ generates linear space of
dimension more that one). Inside $T$ the ground states $\Psi_W$
belong to ${\cal L}_2({\mathbb R}^2)$, $W\in\rm{Int}\ T$. On the
boundary of $T$ the states $\Psi_W$ do not belong to ${\cal
L}_2({\mathbb R}^2)$. Therefore they represent  the bottom of
continuous spectrum for the operators $L_{-},$ and $L^P=L_{+}\oplus
L_{-}$. Apparently, the lowest level for the another sector $L_{+}$,
is strictly {\bf positive.}

Every state $\Psi_W$ for  $L_{-}$ except $c_0$ generate a nonzero
current $J_W$ defined by the phase of the complex function $\psi_W$.
The Current vectors $J_W$ cover some convex bounded domain on the
plane.

{\bf Important fact.} The magnetic flux is divergent on the plane
$R^2$:
$$
\iint\limits_{x^2+y^2\le R^2}Bdxdy= -\frac{1}{2}R\cdot
\oint\limits_0^{2\pi}I_{\{W_j\}}(\varphi) d\varphi+ o(R), \
\text{ïðè} \ R\rightarrow\infty,
$$
where $I_{\{W\}}$ is the indicator of growth (see Appendix).

\noindent II. For the generating functions of the form:
$$
 c=\sum_j\kappa_je^{W_j}+\sum_qe^{W_{R,q}}(\kappa_q'\cos W_{I,q}+\kappa_q''\sin W_{I,q})
$$
we define the indicator of growth using only its part: the real
subset should be chosen $\{W_j\}_+$, such that $\kappa_j>0$. It
defines a domain $T^+$ as above. (Possibly, it is necessary to
change the sign  $c\rightarrow -c$, if this operation leads to the
bigger domain). Anyway, all other linear forms $W_j,W_{R,q}$ should
belong to the domain $T^+$ defined by the main positive part of
linear forms selected above. Under these conditions there exists a
nonempty set of coefficients
  such that $c\ne 0$.  {\bf It is possible to choose  remaining coefficients
  such that  zeroes of $c$  appear}.

\end{enumerate}

\subsection{Solutions of genus g=1. Operators with Bohm-Aharonov Singularity and the Magnetic Bloch functions}

In the current section we study the case of elliptic curves ($g=1$).
As w are going to show,  in this case we are facing a new very
interesting phenomena, connecting out paper with the case of
non-zero magnetic flux \cite{AC,DN,DN2} (magnetic translations and
topological phenomena are also discussed in
\cite{Nov1,Nov2,Lys1,Lys2,Zak}).

Let
$$
 \Gamma'=\Gamma''={\mathbb C}/\Lambda,
$$
where
$$
 \Lambda=\{2m_1\omega_1+2m_2\omega_2,\ m_1,m_2\in{\mathbb Z}\}\subset{\mathbb C}
$$
denotes a lattice. We assume that $\omega_1=1$ and the lattice is invariant with respect
to the complex conjugation
$$
 \bar{\Lambda}=\Lambda
$$
(it is true, in particular, if $\omega_2\in iR$).

Let $\infty_1=0\in\Gamma'$, $\infty_2=0\in\Gamma''$. Assume, that
$Q_0,\dots,Q_n\in\Gamma'$ and $R_0,\dots,R_n\in\Gamma''$ correspond to
the intersection points $ \Gamma'\cap\Gamma''$.

Let us express the  $\psi$-function through the Weierstrass $\sigma$-function
and $\zeta$-function .

We know that $\zeta(w)$ is meromporphic in  ${\mathbb C}$ with first
order poles at the points of $\Lambda$ and
$$
 \zeta(w+2\omega_s)=\zeta(w)+2\eta_s, \eqno{(1)}
$$
where $\eta_s=\zeta(\omega_s).$

The lattice is invariant with respect to the complex conjugation, therefore
$$
 \zeta(w)=\overline{\zeta({\bar{w}})}.
$$

The function $\sigma(w)$ is analytic in ${\mathbb C}$ with first-order
zeroes at the points of $\Lambda$ and
$$
 \sigma(w+2\omega_s)=-e^{2\eta_s(w+\omega_s)}\sigma(w),
$$
$$
 \sigma(w-2\omega_s)=-e^{-2\eta_s(w-\omega_s)}\sigma(w).
$$
The invariance of the lattice with respect to the complex
conjugation implies that
$$
 \sigma(w)=\overline{\sigma({\bar{w}})}.
$$

The function $\psi''=\psi\vert_{\Gamma''}$ has the following form
($z,\bar{z}\in{\mathbb C}, p\in\Gamma'', P=D''$):
$$\psi''(p,z)=
 e^{-z\zeta(p)}\frac{\sigma(p+z+P)}{\sigma(z+P)\sigma(p+P)}.
$$
The function  $\psi'=\psi\vert_{\Gamma'}$ has the  form
$$\psi'(k,z,\bar{z})=
 e^{-\bar{z}\zeta(k)}\left(\frac{\sigma(k+\bar{z}+A_0)\sigma(k-Q_1)\dots\sigma(k-Q_n)}{\sigma(k+P_1)\dots\sigma(k+P_{n+1})}\right.f_0(z,\bar{z})+\dots+
$$
$$
 \left.\frac{\sigma(k+\bar{z}+A_n)\sigma(k-Q_0)\dots\sigma(k-Q_{n-1})}{\sigma(k+P_1)\dots\sigma(k+P_{n+1})}f_n(z,\bar{z})\right),
$$
where
$$
 A_0=Q_1+\dots+Q_n+P_1+\dots+P_{n+1},
$$
$$
 A_1=Q_0+Q_2+\dots+Q_n+P_1+\dots+P_{n+1},
$$
$$
 \dots \dots \dots \dots \dots \dots \dots \dots \dots \dots \dots \dots
$$
$$
 A_n=Q_0+\dots+Q_{n-1}+P_1+\dots+P_{n+1},\ D'=P_1+\dots+P_{n+1}.
$$
The compatibility conditions
$$
 \psi'(Q_s)=\psi''(R_s)
$$
imply
$$
 e^{-\bar{z}\zeta(Q_0)}\frac{\sigma(Q_0+\bar{z}+A_0)\sigma(Q_0-Q_1)\dots\sigma(Q_0-Q_n)}
 {\sigma(Q_0+P_1)\dots\sigma(Q_0+P_{n+1})}f_0(z,\bar{z})=
$$
$$
 e^{-z\zeta(R_0)}\frac{\sigma(R_0+z+P)}{\sigma(z+P)\sigma(R_0+P)},
$$
$$\dots \dots \dots$$
$$
 e^{-\bar{z}\zeta(Q_n)}\frac{\sigma(Q_n+\bar{z}+A_n)\sigma(Q_n-Q_0)\dots\sigma(Q_n-Q_{n-1})}
 {\sigma(Q_n+P_1)\dots\sigma(Q_n+P_{n+1})}f_n(z,\bar{z})=
$$
$$
 e^{-z\zeta(R_n)}\frac{\sigma(R_n+z+P)}{\sigma(z+P)\sigma(R_n+P)},
$$
therefore $f_s(z,\bar{z}):$
$$
 f_0=e^{-z\zeta(R_0)+\bar{z}\zeta(Q_0)}\frac{\sigma(R_0+z+P)}{\sigma(\bar{z}+Q_0+\dots+Q_n+P_1+\dots+P_{n+1})
 \sigma(z+P)}S_0,
$$
$$
 S_0=\frac{\sigma(Q_0+P_1)\dots\sigma(Q_0+P_{n+1})}{\sigma(R_0+P)\sigma(Q_0-Q_1)\dots\sigma(Q_0-Q_n)},
$$
$$
 \dots
$$
$$
 f_n=e^{-z\zeta(R_n)+\bar{z}\zeta(Q_n)}\frac{\sigma(R_n+z+P)}{\sigma(\bar{z}+Q_0+\dots+Q_n+P_1+\dots+P_{n+1})\sigma(z+P)}S_n,
$$
$$
 S_n=\frac{\sigma(Q_n+P_1)\dots\sigma(Q_n+P_{n+1})}{\sigma(R_n+P)\sigma(Q_n-Q_0)\dots\sigma(Q_n-Q_{n-1})}.
$$
We have
$$
 c(z,\bar{z})=
 \left(\frac{\sigma(\bar{z}+A_0)\sigma(-Q_1)\dots\sigma(-Q_n)}{\sigma(P_1)\dots\sigma(P_{n+1})}\right.f_0(z,\bar{z})+\dots+
$$
$$
 \left.\frac{\sigma(\bar{z}+A_n)\sigma(-Q_0)\dots\sigma(-Q_{n-1})}{\sigma(P_1)\dots\sigma(P_{k+1})}f_n(z,\bar{z})\right).
$$
Let us point out that all $f_s$ have the same factor at the denominator
$$
 \sigma'\sigma''=\sigma(\bar{z}+Q_0+\dots+Q_n+P_1+\dots+P_{n+1})\sigma(z+P).
$$
Multiplying all $f_s$ to $\sigma'\sigma''$ we obtain nonsingular
functions $\tilde{c}=c\sigma(\bar{z}+Q+D')\sigma(z+D'')$. The
corresponding functions
$\tilde{\psi}=\psi\sigma(\bar{z}+Q+D')\sigma(z+D'')$ do not have
Bloch-Floquet properties.  They associated with the ground state of
Pauli operator with magnetic field $\tilde{B}=1/2\Delta(\ln
\tilde{c})$. We are going to discuss their analytic properties and
relationship to the  Dubrovin-Novikov  states (1980) below.

Easy to prove following statements.

{\bf Proposition 1:} {\it All solutions $\tilde{c}$ can be presented
in the form
$$\tilde{c}=\sum_q\alpha_q\exp\{-z\zeta(R_q)+\bar{z}\zeta(Q_q)\}\sigma(z+R_q)\sigma(\bar{z}-Q_q)$$
Here $\alpha_q,R_q,Q_q$ can be any generic set of numbers, no
restrictions.}

{\bf Proposition 2:} {\it Every real solution can be presented as a
sum of two type of terms like in Proposition 1: The Type 1 where
$\alpha_q\in \bR$ and $R_q=-\bar{Q}_q$, and The Type 2 where we have
pair of indices $j,l$ satisfying to relations
$\alpha_j=\bar{\alpha}_l, R_j=-\bar{Q}_l, R_l=-\bar{Q}_j$.}

 {\bf
Definition. We say that real solution $\tilde{c}$ has Type $(k,l)$
if it is realized as a sum of $k$ Type 1 terms and $l$ Type 2 terms
(total number of intersection points is $k+2l$).}

 No problem to choose parameters
leading to the nonzero real function $\tilde{c}\neq 0, \tilde{c}\in
\bR$.

{\bf How to choose $\tilde{c}$ with periodic magnetic field $\tilde{B}$ with
same periods as our
lattice?}

{\bf Example.} Let us take $n=2$ (3 intersection points) and real
function $\tilde{c}$ of the type $(1,1)$:

$$\tilde{c}=\alpha\exp\{z\eta+\bar{z}\bar{\eta}\}|\sigma(z-\omega)|^2+$$
$$+\beta\exp\{\bar{z}\zeta(Q_1)-z\zeta(R_1)\}\sigma(z-Q_1)\sigma(\bar{z}+R_1)+
CC$$ Here CC means ''complex conjugate''. We can choose parameters such that this
expression is positive everywhere.

{\bf Proposition 3.} {\it Fix $\alpha\in \bR$ and $
\beta\in \bC$. There
exists a countable number of data $R_2=-R_1=-\bar{Q}_1=\bar{Q}_2,
R_0=-\omega=-Q_0$ such that magnetic field
$\tilde{B}=1/2\Delta\log\tilde{c}$ is  periodic with the same
periods as  lattice. This set is determined by one number which
satisfies to the equations
$$U-\bar{U}=-i\pi n, V+\bar{V}=-\pi m, m,n\in \bZ$$
$$U(\lambda)=\lambda\eta-\omega\zeta(\lambda),
V(\lambda)=\lambda\eta'-\omega'\zeta(\lambda)$$ Sum of such positive
expressions $\tilde{c}=\sum_s\tilde{c}_s$ also leads to the
Algebro-Geometric operator with nonsingular real magnetic field  $\tilde{B}$ with
flux equal to one quantum unit.}

{\bf Conjecture: Every periodic nonsingular magnetic field
$\tilde{B}$ with flux equal to one quantum unit through the
elementary cell, can be approximated
by the Algebro-Geometric Fields described above.}

\vspace{0.3cm}

{\bf Let us make comparison with the Dubrovin-Novikov bases
\cite{DN,DN2}.}

Our function $\tilde{\psi}$ is an eigenfunction for the Pauli
Operator  $L^P$ with the magnetic field $\tilde{B}$. Is it  a
magnetic analog of the Bloch functions? A complete  family of the
Magnetic Bloch Functions, parametrized by the points of the torus
$T^2=\Gamma'$ was constructed in \cite{DN,DN2} for all non-zero
values of magnetic flux for an arbitrary nonsingular periodic field
$\tilde{B(x,y)}$. We consider at the moment the case of flux equal
to one. The whole family determines a direct summand in the Hilbert
space ${\cal L}_2({\mathbb R}^2)$. These functions were written in
the form:
$$\tilde{\psi}=(const)\exp\{S\}\sigma(z+a')$$ where $-1/2\Delta S=\tilde{B}$,
 and  inversion of the Laplacian was
especially normalized. Here $S=\tilde S +(const)z$, where $\tilde S$ is a real
functions, independent on $a'$ and $(const)$ depends on $a'$ and the lattice.
In our case we start with function $\psi''$
holomorphic in $z$ because the actual ground states (corresponding
to the spectrum in  the Hilbert Space) belong to the spin-sector
described by the curve $\Gamma''$:
 We have  to reduce the operator $L^P=L^+\bigoplus L^-$ to the
self-adjoint form by the ''Gauge Transformation''.   We  choose the
proper spin-sector where the ground states are located, i.e.
eigenfunctions decay at infinity (as for genus zero). It corresponds
to the curve $\Gamma''$. For that sector we need to divide
$\psi=\psi''$ by the $\sqrt{c}$. Use here notations $a=P, b=A, d=p-P$.
So we have finally
$$\psi(p,z)=(1/\sqrt{c} )\psi''(p,z)
=(\sqrt{\sigma(z-a)\sigma(\bar{z}-b)})/(\sqrt{\tilde{c}})e^{-\zeta(d)z}
\frac{\sigma(z+d-a)}{\sigma(z-a)\sigma(d-a)}=$$

$$=(const)\sqrt{\frac{\sigma(\bar{z}-b)}{\sigma(z-a)}} (1/\sqrt{\tilde{c}})e^{-\zeta(d)z}\sigma(z+d-a)$$
where $p\in \Gamma''=C/\Lambda$. In order to remove a singular
Bohm-Aharonov $\delta$-term we apply a unitary singular gauge
transformation and multiply the result to a linear exponent
$$\psi\rightarrow\tilde{\psi}=\sqrt{\frac{\sigma(z-a)}{\sigma(\bar{z}-b)}}
e^{(const) z}\psi
$$
where $(const)$ depending on $p$ is chosen to obtain bounded magnetic Bloch eigenfunctions.

The last function $\tilde{\psi}$ leads exactly to the
Dubrovin-Novikov magnetic Bloch  family of eigenfunctions for the smooth
field $\tilde{B}$:
$$\tilde{\psi}=(const)(1/\sqrt{\tilde{c}})e^{-\zeta(d)z}e^{(const) z}\sigma(z+d-a)=\exp\{S\}\sigma(z+a')$$
For smooth fields this family is uniquely defined.

Another form of this argument is following: we can always multiply
$\psi''$ by any holomorphic function, which we choose as $\sigma(z-a)e^{(const) z}$
with cons depending on $p$.
The product also satisfies to the Cauchy-Riemann (=''self-duality'')
equation as an ''Instanton'' in the first nonself-adjoint form of
the Pauli operator. After that we multiply result by the
$1/\sqrt{\tilde{c}}$ realizing a non unitary ''Gauge
Transformation'' leading to the self-adjoint form of $L^P$ in the
sector where ground states are located. Finally we get exactly the
Dubrovin-Novikov Magnetic Bloch function (see{\cite{DN}).

These arguments work because we are dealing with the ''Instanton
Family'' (i.e. satisfying to the first order Cauchy-Riemann equation
$\bar{\partial} \psi''=0$)  for the operator $L^P$ written in the
first non-selfadjoint form. It satisfies to the covariant first
order equation $[\bar{\partial}+1/2\partial (\ln
\tilde{c})][(1/\sqrt{\tilde{c}})\psi'']=0 $ after nonunitary gauge
transformation realized  as a division of all eigenfunctions
including ground-states by the factor $\sqrt{\tilde{c}}$. Here
$a,b,d$ are constants with the proper reality restrictions replacing
the points $a=P,b=A=\sum_sP_s-\sum_sQ_s,d=p-P$ correspondingly on
the elliptic curve. In particular $b=\bar a$.

{\bf For  $g>1$ we always have magnetic flux with more than one
quantum unit. We have to use Riemann surfaces $\Gamma'$ with
selected point $\infty_1$ such that solutions of  the corresponding
KP hierarchy is Elliptic in the variable $x$ in order to get
magnetic fields $B$ and $\tilde{B}$ periodic in both directions in
the $z$-plane.} The theory of elliptic solutions to the KdV equation
was started
 by Dubrovin and Novikov in 1974. For the KP hierarchy it was developed
 by Krichever since 1979 . A number
  of works were dedicated to it in the later literature.
  Details will be presented in the next work of the present
authors.

\vspace{0.3cm}

{\bf Let us discuss here an extremely important physical question:
What is a quantum Bohm-Aharonov Phenomenon? How the $\delta$-term
in magnetic field affects the spectrum?}

\vspace{0.3cm}

In our case the singular magnetic field $B$ with the $\delta$-type singularity
has  an algebro-geometric realization. It has a   zero magnetic flux.
The family of the complex Bloch-Floquet eigenfunctions is found for it which
has very specific
 analytical properties
  valid only in the case of zero total flux through the elementary cell. It
was explicitly calculated
in the appropriate spin-sector. Its ''instanton part'' is simply the Baker-Akhiezer
function $\psi''$. After  reduction to the self-adjoint form of operators $L^P$
it became $\psi''/\sqrt{c}$ with same quasimomentum. Other part $\psi'/\sqrt{c}$ certainly
does not have the instanton form outside of the intersection points.
  Is everything correctly and uniquely defined
for the singular operators of this type? Are the Bloch-Floquet multiplicators (whose logarithms
divided by periods define the components of quasimomentum)
canonically well-defined for such singular operators?

This question needs clarification. {\bf Our statement is following: This family
 is correctly defined
as a limit of corresponding families for the smooth operators with zero magnetic flux. }
Such a procedure to define
spectrum and the whole complex  family of Bloch-Floquet functions
can be realized for the zero level by the family of Riemann surfaces $\Gamma_{\tau}$
degenerating to
 $\Gamma_0=\Gamma'\bigcap\Gamma''$.
 As we can see,
only some special isolated state in this family might serve the standard Hilbert Space.
In the Hilbert space $\cL_2(\bR^2)$ such state corresponds to the
bottom of the continuous spectrum for the field $B$. So either the spectrum near 0 is
 continuous or the dispersion relation near the zero point is identically trivial, and
 we have in fact more Bloch functions on the level $\epsilon=0$.
 It is exactly the case here. We will clarify this question below.

 The Bloch-Floquet multiplicators of the family $\psi''/\sqrt{c}$ are $\kappa_x,\kappa_y$
 They  are equal to
 $$\kappa_x=\exp\{-2\omega\zeta(p)-2\eta p\},
\kappa_y=\exp\{-2i\zeta(p)\omega'-2i\eta' p\}$$
The equations $|\kappa_x|=1,|\kappa_y|=1$ can be easily solved
$$p^R=-\zeta(p)^R\omega/\eta,p^I=-\zeta^I\omega'/eta'$$
or $$-\zeta(p)=p^R\eta/\omega+ip^I\eta'/\omega'$$
So they are non-unitary
$|\kappa|\neq 1$ for other points $p\in \Gamma''$.

 {\bf Removing singular part from
$B$ we are coming to the  Magnetic-Bloch Functions found in 1980: the
multiplicators (i.e.the eigenvalues of the Magnetic Translations)
became unimodular $|\tilde{\kappa}|=1$. Remaining spectrum will be
separated from 0 by the finite gap for the field $\tilde{B}$.}

The formal procedure  is following: We introduce function
$$\psi_{new}(p,z)=(\psi''/\sqrt{c})\exp\{u(p)z\}$$
choosing $u(p)$ such that all multiplicators became unitary
$\kappa\rightarrow\tilde{\kappa}$:
$$|\kappa_xe^{2u(p)\omega}|=1=|\kappa_ye^{2iu(p)\omega'}|$$
For the new multiplicators (after this renormalization of eigenfunction)
we have $$\tilde{\kappa}_x=\exp\{2ip^I[\eta'\omega-\eta\omega']/\omega'\},
\tilde{\kappa}_y=\exp\{2ip^R[\eta\omega'-\eta'\omega]/\omega\}$$
Finally we remove  singularity by multiplication $\psi_{new}\rightarrow |\sigma'|\psi_{new}$
 replacing $c$ by the smooth $\tilde{c}$ in the formula above.
So we are coming exactly to the magnetic Bloch eigenfunctions of Dubrovin and Novikov(1980)
with  multiplicators  $\tilde{\kappa}_x,\tilde{\kappa}_y$
(may be shifted by constant which is inessential).

The multiplicators $\tilde{\kappa}$
form together a point of the 2-torus $\tilde{\kappa}\in T^2$ forming the whole component
space of the real
  quasimomentum  at the zero level and nearby. We have  a big complex 2-dimensional manifold $M^2$
  of the Bloch-Floquet eigenfunctions for the singular operator
 $L^P$ with zero flux
of the form $$M^2=\Gamma''\times CP^1$$ compactified at the infinities
 $u\in CP^1=C\bigcup \infty$.
It presents an irreducible component
 of the whole Bloch-Floquet
manifold for this operator.  The  dispersion relation
 $\epsilon\rightarrow C$ degenerates at this component, i.e.
 $\epsilon=0$ identically. The Bloch eigenfunction already written above has a form
$$\Psi(p,u,z)=\psi''/\sqrt{c}\times\exp\{uz\}$$ where $(p,u)\in M^2$.
This manifold presents exactly one component of the limit of the whole Bloch-Floquet
 manifolds $M^2_{\tau}$ for $\tau\rightarrow 0$ and $M^2_0=M^2\bigcup M'$.
Every small purely magnetic perturbation leading to the smooth magnetic field $B_{\tau}$
closed to our singular field $B$, has quite similar Bloch-Floquet eigenfunction:
 The Bloch manifold here is
 $\Gamma''_{\tau}\times CP^1$ with the instanton-type  Bloch eigenfunction like
$\Psi_{\tau}=\psi''_{\tau}(p,z)/\sqrt{c_{\tau}}\times \exp\{uz\}$ where the curve $\Gamma''_{\tau}$ might
have an infinite genus. So the small electric perturbation of operator $L^P$
is needed. We perturb by  the small
 periodic potential $\tau U(x,y)$. Already the first order in coupling parameter $\tau$
 probably leads to a nontrivial
  dispersion relation as a complex meromorphis function on the same manifold.  It has
   the order $\tau$, so we have
  $$\epsilon_{\tau}:M^2_{\tau}\rightarrow C$$
  The real levels $\epsilon_{\tau}=const\in R$  give a function on the real
  ''quasimomentum'' torus $T^2$ whose levels are the real Fermi-curves.
   Its minimum lies nearby of the initial point on the curve $\Gamma''$ found above
    where $\kappa_x,\kappa_y$ are
   imaginary.

  We are going to
  calculate this perturbation in the next work. The complex level curves should have
  analytical properties at infinity typical for the Baker-Akhiezer functions (
  maybe of the infinite genus
  where small handles appear from the ''resonance points'').

 So it looks like the delta-term does not affect deeply the spectrum near the ground state.

\section{Appendix: The Asymptotic of Magnetic Flux}

Let us calculate the Asymptotic of Magnetic Flux through the round ball of radius $R$ for the purely
exponential case
 $$e^{2\Phi}=c=\sum_j\kappa_je^{W_j},\kappa_j>0,W_j=R(\alpha_j\cos(\phi)+\beta_j\sin(\phi))$$

We have for the magnetic field $B=-(\Delta(\ln c))/2$. For the vector-potential restricted on the circle
$r=R$ in polar coordinates, we obtain
$$
A=\Phi_ydx-\Phi_xdy=-\frac{1}{2}R[\sum_j\kappa_je^{W_j}(\alpha_j\cos(\phi)+b_j\sin(\phi))]d\phi/c.
$$
Our assumption is that there exist exactly $N$

indices $j=1,2,...,N$ such that the Indicator of our family
$I_{\{W_j\}}(\phi)=max_jI_{W_j}(\phi)$ where
$I_{W_j}=max[\alpha_j\cos(\phi)+\beta_j\sin(\phi),0]$ is strictly positive, and all other indices $p\neq 1,2,...N$
are inessential (i.e. corresponding linear forms $W_p$ are located strictly inside of the convex domain
$T\subset R^2$ with coordinates $\alpha,\beta$ numerating the rapidly decreasing ground state vectors of our operator).
There are domains $\Delta_j$ on the circle $S^1$, where $I_{\{W_k\}}=I_{W_j}(\phi)$ with end points
$\Delta_j=[\phi^0_j,\phi^0_{j-1}]$, and for $j=N,1$ we have $\phi_0^0=\phi^0_N$. So $\Delta_N$ is a neighbor of $\Delta_{N-1}$ and
$\Delta_1$ (i.e. our numeration ic circle contr-clockwise).

Our claim is following:
{\bf Following Asymptotic Formula is true:}

$$\int\int_{D^2_R}B(x,y)dxdy+\frac{1}{2}R\oint_{S^1}I_{\{W_k\}}(\phi) d\phi=$$
$$
 =\sum_{s\geq 1}R^{-s}\sum_{j=1}^N\lambda^{(s)}_j\{Q_s(a_j)+Q_s(a_j^{-1})(-1)^s\}
+ \text{(Remainder)}.
$$
Apparently, this series is nonconvergent, since the coefficients $Q_s$ grow, as we think, as  $s!$.
About the remainder we claim now that its decay is more rapid than any negative degree of $R$.
We claim only that the
{\bf ''Regularized Flux''}
$$
\int\int_{D^2_R}Bdxdy+\frac{1}{2}R\oint_{S^1_R}I_{\{W_j\}}(\phi)d\phi=O\left(\frac{1}{R}\right)
$$
{\bf is tend to zero in this sum, for  $R\rightarrow\infty$.} Performing this calculation near the critical points $\phi_j^0$, we use following functions
$$
 (W_{j+1}-W_j)/R=(\alpha_{j+1}-\alpha_k)\cos(\phi)+(\beta_{j+1}-\beta_j)\sin(\phi)=t_j(z).
$$
Here $\phi=(\phi_j^0+ z), |z|<\epsilon$. It is located near the points $\phi^0_j$ or $z=0$: in this point $W_j=W_{j+1},t_j=0,z=0$, and
the inverse function $z(t_j))$ is given by the inverse series with a finite radius:
$$z=\sum_{k\geq 1}\lambda^{(k)}_jt_j^{k+1}/(k+1)$$
$$d\phi=dz=\sum_{k\geq 0}\lambda^{(k)}_jt_j^kdt_j$$

We define numbers
$$Q_k(a)=\int_0^{\infty}[aw^ke^{-w}/(1+ae^{-w})]dw$$
useful for the investigation of the difference
$$
 \oint_{S^1_R}A+\frac{1}{2}R\oint_{S^1}I_{\{W_q\}}(\phi)d\phi.
$$
Probably, $Q_k\thicksim k!$.
Our function $c$ has exponential growth everywhere, but magnetic field has decay only outside of the small domains surrounding
the ''critical'' points $\phi^0_j$. It is easy to see that our vector-potential $A$ after extracting the Indicator of Growth
$RI_{\{W_q\}}(\phi)(d\phi)$, became exponentially small outside these small domains. Only two exponential terms
$\kappa_je^{W_j},$ $\kappa_{j+1}e^{W_{j+1}}$ in $c$ are essential in every such small domain, between $\Delta_j$ and $\Delta_{j+1}$.
Dropping all other terms in the sum for $c=\sum_q\kappa_qe^{W_q}$ and for $A$ in every such small area
costs us exponentially small. In the area $\phi\in \Delta_q$ we multiply both-numerator and denominator in the expression for $A$-by
the exponent $\kappa_q^{-1}e^{-W_q}$. The exponent $e^{-W_j}$ is the vertex of the convex polygon $T$
containing all functions $c'=ce^W\in T$ such that $(ce^{W_q})^{-1/2}$ are the ground states of the Pauli Operator.
We need $q=j$ for $\phi\leq \phi_j^0$ (or $\phi\in\Delta_j$), and $q=j+1$ for $\phi\in\Delta_{j+1}$.
So only two terms remain in the numerator and denominator.

Similar result we obtain in the domain $\Delta_j$ just below the point $\phi_j^0$ with inverse constant $\kappa_{j+1}/\kappa_j$
and exponent $e^{\{ W_{j+1}-W_j\}}$, plus we have to turn back the direction of integration.
Taking $\epsilon$ such that $R\epsilon=O(R^{\delta}),\delta>0,$ we see following:
The integration between the local limits $[\phi_j^0-\epsilon,\phi^0_j+\epsilon]$ of such expressions with
$w=Rt_j$, which appear in our calculation of the regularized magnetic flux, can be extended to the limits $[-\infty,+\infty].$
It is true because the remaining terms have order $O(e^{-R^{\delta}})$: more precisely their decay is more rapid than any polynomial.

Expressing the variable $z=\phi-\phi^0_j$ by the variable $t_j=(W_j-W_{j+1})/R$, we are easily coming to our result.
In the final integration we have a sum of integrals looking like
$$Q_s(a)=R^{-s-1}\int_0^{\infty}ae^{-w}/(1+ae^{-w})w^sdw,$$
where
$w=\pm Rt_j$.  The sign is $+$ and $a=\kappa_{j+1}/\kappa_j$ for $z\leq 0$, and sign
$-$ and $a$ replaced by $a^{-1}$ for $z\geq 0$. So we are coming to our result.

Note, it is easy to show that the expressions
$$
 Q_k'=\int_0^{\infty}e^{-w}w^kdw
$$
 grow as $(k!).$  Probably, it is true for our expressions $Q_k$.

\end{document}